\documentclass[twocolumn,showpacs,preprintnumbers,amsmath,amssymb]{revtex4}
\usepackage{graphicx}
\usepackage[dvipsnames]{color}
\usepackage{bm}
\begin{document}
\def \brho{{\hbox{\boldmath $\rho$}}}
\def \beps{{\hbox{\boldmath $\epsilon$}}}
\def \bdelta{{\hbox{\boldmath $\delta$}}}
\title{Quantum friction between oscillating crystal slabs: Graphene monolayers on dielectric substrates}

\author{Vito Despoja$^{1,2,3}$}
\author{Pedro M. Echenique$^{1,2}$}
\author{Marijan \v Sunji\' c$^{1,4}$}
\affiliation{$^1$Donostia International Physics Center (DIPC), P. Manuel de Lardizabal, 20018 San Sebastian, Basque Country, Spain}
\affiliation{$^2$Departamento de Fisica de Materiales and Centro Mixto CSIC-UPV/EHU, Facultad de Ciencias Quimicas, Universidad 
del Pais Vasco UPV/EHU, Apto. 1072, 20080 San Sebastian, Basque Country, Spain}
\affiliation{$^3$Institute of Physics, Bijeni\v cka 46, HR-10000 Zagreb, Croatia}
\affiliation{$^4$Department of Physics, Faculty of Science, University of Zagreb, Bijeni\v{c}ka 32, HR-10000 Zagreb, Croatia}

\begin{abstract}
We present a theoretical description of energy transfer 
processes between two noncontact quasi-twodimensional crystals separated by distance $a$, 
oscillating with frequency $\omega_0$ and amplitude $\rho_0$, and compare it with the case of two quasi-twodimensional crystals in uniform 
parallel motion. We apply the theory to calculate van der Waals energy and dissipated energy in two oscillating slabs where each slab 
consists of a graphene monolayer deposited on SiO$_2$ substrate. The graphene dielectric 
response is determined from first principles, and SiO$_2$ surface response is described using 
empirical local dielectric function.    
We studied the modification of vdW attraction  as function of the
driving frequency and graphene doping.
We propose the idea of controlling the 'sticking' and 'unsticking' of slabs by tuning the graphene 
dopings $E_{Fi}$ and driving frequency $\omega_0$.
We found simple $\rho^2_0$ dependence of vdW and dissipated energy. 
As the Dirac plasmons  are the dominant channels through which the energy between slabs can be 
transferred, the dissipated power in equally doped  $E_{F1}=E_{F2}\ne 0$ graphenes shows strong 
$\omega_0=2\omega_p$ peak. This peak is substantially reduceed when graphenes are deposited on SiO$_2$ substrate.   
If only one graphene is pristine ($E_{Fi}=0$) the $2\omega_p$ peak disappears.  
For larger separations $a$ the phononic losses also become important and the doping causes shifts, appearance  and disappearance of 
many peaks originating from resonant coupling between hybridized electronic/phononic excitations in graphene/substrate slabs.
\end{abstract}
\maketitle
\section{Introduction}
Detailed understanding of non-contact friction and energy transfer processes in nanostructures is of great importance, both 
from the conceptual and practical viewpoints.
Existing theoretical studies, starting with the seminal paper by Pendry \cite{Pendry1}, mostly consist of calculations of friction coefficients, i.e. friction force between two parallel dielectric plates (e.g. supported graphenes) in uniform relative motion which is experimentally not easily measured (e.g. current drag in one graphene caused by  current flow in another 
one) \cite{Theory1,Theory2,Theory3,Theory4,Theory5,Theory6,Theory7}.

While the experiments with two slabs in parallel relative motion with constant velocity are difficult to perform, we suggest here 
that for the same systems experiments with slabs in relative oscillatory motion with fixed or variable frequency might be easier to 
perform, and could lead to new and interesting observations. 
Recently a similar approach has been realized experimentally \cite{QF_exp_osc1,QF_exp_theory_osc,QF_exp_osc2,QF_exp_osc3}.
In these experiments the system (usually an AFM tip above the surface) oscillates at some characteristic 
frequency. These oscillations are then, because of various dissipation mechanisms (which includes quantum friction), damped.       
Our model is based on a slightly different concept; one of the slabs, e.g. the AFM tip, is driven with variable frequency. This means that the friction can 
be deduced from the energy dissipated in one oscillating cycle. 
In this paper we provide a general theoretical description of such processes, expecting that this method might become a useful tool to study dynamical properties of 
low-dimensional systems \cite{Munez}.

The main objective of this paper is therefore a theoretical description of these phenomena in systems consisting 
of two non-touching polarizable media, specifically conservative (van der Waals or Casimir) and 
dissipative forces (quantum friction) between two quasi-twodimensional (q2D crystals) in 
relative parallel and oscillatory motion. While the case of slabs in parallel uniform motion has been extensively studied \cite{Pendry1,Persson1,Volokitin1,
Volokitin2,Theory4,Pendry2,Philbin}, here we develop an analogous theory describing interaction of atomically thick slabs (q2D crystals) in oscillatory motion.

In Sec.\ref{Sec2} the expressions for van der Waals and dissipative energies and forces are derived for such a q2D system in a very general case,
for variable slab temperatures and dynamical properties characterized by their surface response functions $D_1$ and 
$D_2$, and for variable oscillating frequencies and amplitudes. 
We assume 2D translational invariance and neglect retardation for the slab distances in consideration. 
For the sake of clarity and comparison, in Appendix\ref{AppA} we derive analogous results for the 
case of parallel uniform motion, recovering but also generalizing some earlier results \cite{trenjePRB,PFK}.

In Sec.\ref{DerofD} we derive general expressions for surface response functions $D_i$ for multilayer slabs, later to 
be specified  for monolayers of a substance like graphene or silicene adsorbed on dielectric substrates. Surface response functions 
$D_1$ and $D_2$ will be the key ingredients in the expressions describing dissipative and 
reactive processes in Sec.\ref{Sec2} and Sec.\ref{DerofD}.  
In Sec.\ref{DerofD} we also show how to calculate surface response functions $D_i$ for a specific case of q2D crystals on a dielectric substrate
The expression for the surface excitation propagator of a system of 
two coupled slabs is also derived.

In Sec.\ref{nabnak} we present the models used to describe the q2D crystal  and substrate dynamical response. 
We study the specific case of a graphene monolayer on a dielectric substrate, which is chosen to be ionic  
crystal SiO$_2$. 
The substrate is considered as a homogenous semiinfinite ionic crystal SiO$_2$ with the appropriate dielectric function in 
the longwavelength limit. Graphene monolayer dynamical  response is determined from 
first principles. Also some computational details are specified.

In Sec.\ref{resuuult}  general expressions of previous sections are applied to the system of 
two slabs, where each slab represents a graphene($E_{Fi}$)/SiO$_2$ system, and where graphene doping is characterized 
by Fermi energy $E_F$ relative to the Dirac point.

In Sec.\ref{cmspe} we demonstrate how  the spectra of electronic excitations in one slab 
and in two coupled slabs depend on graphene doping $E_F$.

The form of these coupled ecitations is responsible for the behaviour of the atractive forces and dissipation.
We first discuss in Sec.\ref{weakvdW} the modification of van der Waals force for oscillating in comparison 
with the static slabs. Van der Waals energies depend on two factors. They increase with the increased 
graphene doping, but are reduced for  the asymmetric doping when excitations in two slabs are off-resonance. 
Dynamical vdW energy shows unusual behavior: it starts as  plateau,  and then decreases.
This is, because the fast Dirac plasmon in one slab for low driving frequencies $\omega_0<\omega_p$, still  perfectly follows Doppler shifted  charge density fluctuations in another slab. 
For larger driving frequencies this is not the case and vdW energies decrease.    
Finally,  for small or zero doping the $\pi\rightarrow\pi^*$  and $\pi\rightarrow\sigma$ excitations 
cause linear weakening of the dynamical vdW energy.   
       
In Sec.\ref{DISsub} we calculate and discuss how dissipated power depends on various parameters: 
driving amplitude $\rho_0$ and frequency  $\omega_0$, on the separations between slabs $a$ and on the 
substrate. We find simple $\rho^2_0$ dependence, while the $\omega_0$ dependence is determined 
by the intensity of resonant coupling between hybridized Dirac plasmons and substrate TO phonons.
We found that in realistic grahenes (in comparison with Drude model when excitation of undamped Dirac plasmons provides 
unrealistically strong $2\omega_p$ peak in the dissipated power)  the  dissipation power peak is strongly reduced and  red shifted.
We also explain why the substrate substantially reduces dissipated power peak.  
For larger separations $a$ additional peaks appear in dissipated power originating from the excitations of hybridized substrate phonons.     

In Sec.\ref{DISdop} we explore how the dissipated power depends on graphene dopings.
We show that if one graphene is pristine ($E_F=0$) it causes the disappearance of strong $2\omega_p$ peak 
in the dissipated power. Moreover, for larger separations the doping causes shifts, appearance  and disappearance  of many peaks originating 
from resonant coupling between hybridized substrate phonons and Dirac plasmons.

In Sec.\ref{Sec5} we present the conclusions.

\section{General theory: Oscillating slabs}
\label{Sec2}

\subsection{Van der Waals energy and force}
In Appendix \ref{vdWenergyforce} we have derived van der Waals energy and force between two slabs 
in uniform relative motion in some detail because it will help us to treat a 
similar problem of two oscillating slabs. 

\begin{figure}[t]
\includegraphics[width=0.9\columnwidth]{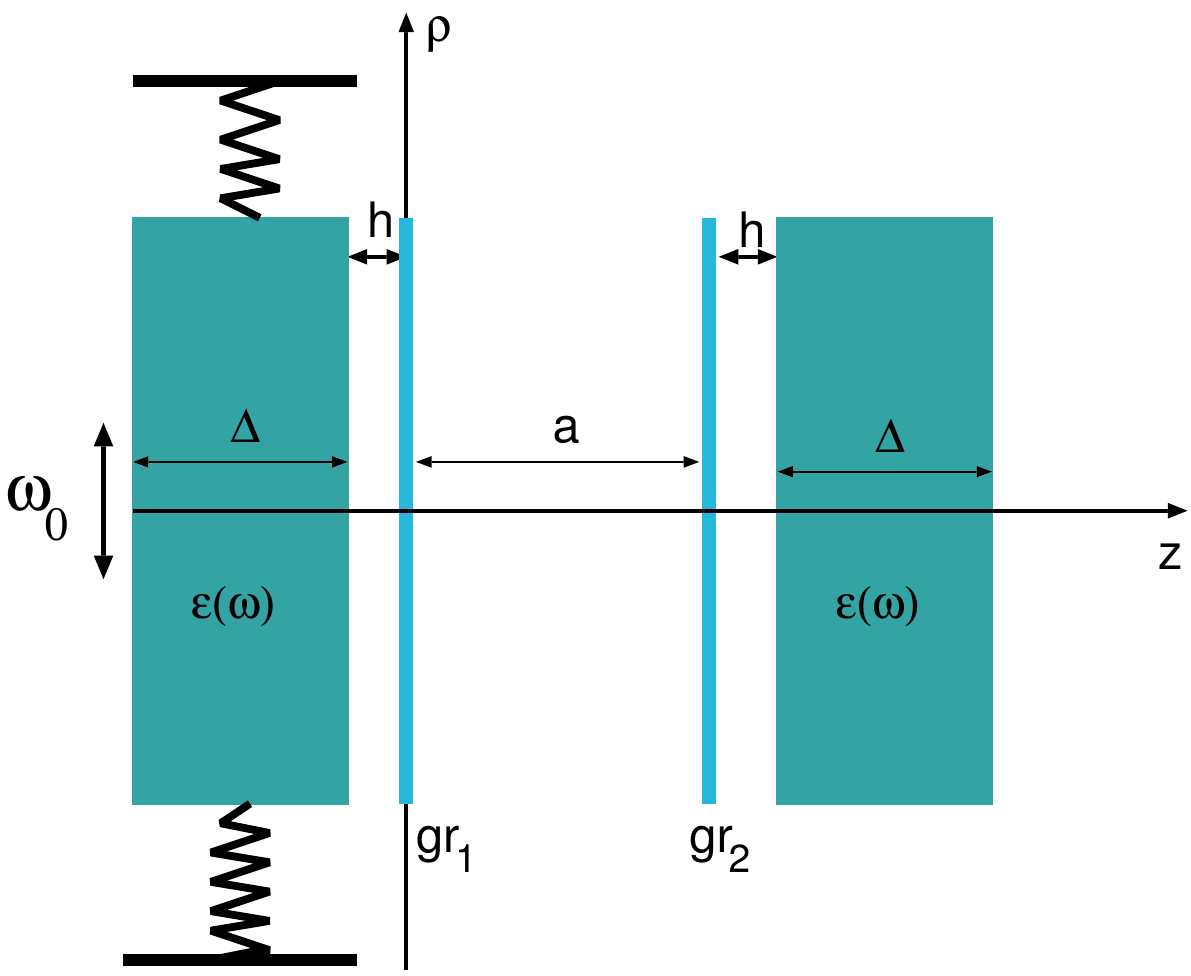}
\caption{Geometry of the system.}
\label{Fig1}
\end{figure}

We shall later assume that the slabs consist of graphene monolayers with variable doping, deposited on dielectric slabs of thickness $\Delta$ 
described by local  dielectric functions $\epsilon(\omega)$, as shown in Fig.\ref{Fig1}. The left slab mechanically oscillates with frequency $\omega_0$  and 
amplitude $\brho_0$ relative to the right slab. Again we calculate the diagram in Fig.\ref{FigA1} as in the \ref{vdWenergyforce}, but now  the slab parallel coordinates change in time 
as 
\begin{equation}
\brho-\brho_1\rightarrow\brho-\brho_1-\brho_0(\sin\omega_0t-\sin\omega_0t_1)
\label{kukuli}
\end{equation}
so that instead of (\ref{nabij3}) we have
\begin{equation}
\begin{array}{c}
E_{c}=\int^{\infty}_{-\infty}dt_1\int\frac{d{\bf Q}}{(2\pi)^2}
e^{-i{\bf Q}\brho_0(\sin\omega_0t-\sin\omega_0t_1)}
\nonumber\\
\nonumber\\
\int^{\infty}_{-\infty} dzdz_1dz_2dz_3 
S_1({\bf Q},z,z_1,t-t_1)V({\bf Q},z,z_3)
\nonumber\\
\nonumber\\
D_2({\bf Q},z_3,z_2,t-t_1)V({\bf Q},z_2,z_1).  
\end{array}
\label{bijem3}
\end{equation}

If we use 
\[
e^{iz\sin\phi}=\sum^{\infty}_{m=-\infty}J_m(z)e^{im\phi} 
\]
where $J_m$ are Bessel functions, after Fourier transformation in $\omega$ space, using expressions (\ref{Def1}--\ref{Def3}), (\ref{impsv}) and 
integration over $z$ coordinates we obtain  
\begin{equation}
\begin{array}{c}
E_{c}=\hbar\int\frac{d{\bf Q}}{(2\pi)^2}e^{-2Qa}
\sum^{\infty}_{m,m'=-\infty}J_m({\bf Q}\brho_0)J_{m'}({\bf Q}\brho_0)\hspace{3cm}
\nonumber\\
\nonumber\\
\int^{\infty}_{-\infty}\frac{d\omega}{2\pi}
\left[2n_1(\omega)+1\right]e^{i(m-m')\omega_0t}\hspace{3cm}
\nonumber\\
\nonumber\\
ImD_1({\bf Q},\omega)ReD_2({\bf Q},\omega+m\omega_0).\hspace{3cm}
\end{array}
\end{equation}
Here we have also used the fact that $ImD_2({\bf Q},\omega)$ is an antisymmetric function of 
$\omega$ and does not contribute to integration. 
We see that the energy  oscillates in time with frequencies $(m-m')\omega_0$. 
If we assume to measure energies on a time scale $\Delta t>T$, where $T=\frac{2\pi}{\omega_0}$ is the maximal 
duration of one cycle, then we can average over $T$
\begin{equation}
\frac{1}{T}\int^{T}_0dte^{i(m-m')\omega_0t}=\delta_{mm'},
\label{Timeaverage}
\end{equation}
and find the result independent of time: 
\[
\begin{array}{c}
E_{c}=\frac{\hbar}{2}\int\frac{d{\bf Q}}{(2\pi)^2}e^{-2Qa}
\sum^{\infty}_{m=0}(2-\delta_{m0})J^2_m({\bf Q}\brho_0)
\nonumber\\
\nonumber\\
\int^{\infty}_{-\infty}\frac{d\omega}{2\pi}\ \left\{[2n_1(\omega)+1]
ImD_1({\bf Q},\omega)ReD_2({\bf Q},\omega+m\omega_0)+\right.
\nonumber\\
\nonumber\\
\left.[2n_2(\omega)+1]ImD_2({\bf Q},\omega)ReD_1({\bf Q},\omega+m\omega_0)\right\},
\end{array}
\]
where the expression in curly brackets is fully analogous to the one in (\ref{jura66}), but now $\omega'\rightarrow\omega_m=\omega+m\omega_0$.
Inclusion of higher order processes follows the same procedure as for the parallel motion in \ref{vdWenergyforce}. After 
integration over the coupling constant, we obtain the result analogous to (\ref{jujujuju7}) 
\begin{eqnarray}
E_{c}=\frac{\hbar}{2}\int\frac{d{\bf Q}}{(2\pi)^2}\sum^{\infty}_{m=0}(2-\delta_{m0})J^2_m({\bf Q}\brho_0)\times
\label{vdwFIN}\\
\nonumber\\
\int^{\infty}_{-\infty}\frac{d\omega}{2\pi}A({\bf Q},\omega,\omega_m)\hspace{3cm}
\nonumber
\end{eqnarray}
where $A$ is given by (\ref{jalko}) and (\ref{jujucka}), with $\omega_m=\omega+m\omega_0$.

Again, the limiting cases can be obtained from Sec.\ref{vdWenergyforce}.
For $\omega_0=0\ (\omega'=\omega)$ and $\brho_0=0$ we find the well known result for van der 
Waals interaction when the slabs are at rest \cite{vdW2007,Pedro}:    
\begin{eqnarray}
E_{c}(a)=\frac{\hbar}{2}\int\frac{d{\bf Q}}{(2\pi)^2}\int^{\infty}_{0}\frac{d\omega}{2\pi}\ sgn\omega\times\hspace{2cm} 
\nonumber\\
\nonumber\\
\hspace{3cm}Im\ln\left[1-e^{-2Qa}D_1({\bf Q},\omega)D_2({\bf Q},\omega)\right]
\nonumber
\end{eqnarray}
For finite frequency $\omega_0$ and $D_1=D_2=D$ we find: 
\begin{eqnarray}
E_{c}(a)=\frac{\hbar}{2}\int\frac{d{\bf Q}}{(2\pi)^2}
\sum^{\infty}_{m=0}(2-\delta_{m0})J^2_m({\bf Q}\brho_0)
\nonumber\\
\nonumber\\
\int^{\infty}_{-\infty}\frac{d\omega}{2\pi}sgn\omega\ 
Im \ln\left[1-e^{-2Qa}D({\bf Q},\omega)D({\bf Q},\omega_m)\right].
\nonumber
\end{eqnarray}
We notice that the frequency integrals are the same as 
in (\ref{jujujuju7}--\ref{sinko}). Also, the attractive van der Waals force 
between two oscillating slabs is given by 
\begin{eqnarray}
F_{\perp}(a)=-\frac{dE_{c}(a)}{da}=\hspace{3cm}
\nonumber\\
\nonumber\\
\hbar\int\frac{d{\bf Q}}{(2\pi)^2}Qe^{-2Qa}
\sum^{\infty}_{n=0}(2-\delta_{m0})J^2_m({\bf Q}\brho_0)\times
\nonumber\\
\nonumber\\
\int^{\infty}_{-\infty}\frac{d\omega}{2\pi}B({\bf Q},\omega,\omega_m) 
\end{eqnarray}
where the function $B$ is given by (\ref{prcko}) and (\ref{prdf99}).   
The same holds for the $\omega_0\rightarrow 0$ or $D_1=D_2=D$ limits when the 
expressions for $B$ become (\ref{jurec}) or (\ref{kurec}), respectively.
\subsection{Dissipated power}
We can perform the calculation of the dissipated power for two slabs oscillating parallel to each 
other with amplitude $\brho_0$ and frquency $\omega_0$ in analogy with the previous treatment of 
two slabs in uniform relative motion in Sec.\ref{jurniga}. Again, we have to transform the parallel 
coordinates in the left slabs as in (\ref{kukuli}).
Then (\ref{losse5}), after integration over $t_1$ becomes 
\begin{equation}
\begin{array}{c}
P_{12}(t)=-i\hbar\int\frac{d{\bf Q}}{(2\pi)^2}\int\frac{d\omega}{2\pi}\sum^{\infty}_{m,m'=-\infty}
\\
\\
(-1)^{m+m'}e^{i(m'-m)\omega_0t}(m'\omega_0-\omega)\ 
J_m({\bf Q}\brho_0)J_m'({\bf Q}\brho_0)
\\
\\
S_1({\bf Q},|\omega|,z,z_1)\otimes V({\bf Q},z,z_3)\otimes 
\\
\\
D_2({\bf Q},m'\omega_0-\omega,z_3,z_2)
\otimes V({\bf Q},z_2,z_1) 
\end{array}
\end{equation}
We see that the energy transfer rate is time dependent and oscillates 
with frequency $(m'-m)\omega_0$. Again, from (\ref{Timeaverage}) we see 
that for time intervals large with respect to the oscillation period $T$ the 
terms $m\ne m'$ do not contribute and the energy transfer rate is
\begin{equation}
\begin{array}{c}
P_{12}=
-i\hbar\int\frac{d{\bf Q}}{(2\pi)^2}\int\frac{d\omega}{2\pi}\sum^{\infty}_{m=-\infty}(m\omega_0-\omega)\ 
J^2_m({\bf Q}\brho_0)
\\
\\
S_1({\bf Q},|\omega|,z,z_1)\otimes V({\bf Q},z,z_3)\otimes 
\\
\\
D_2({\bf Q},m\omega_0-\omega,z_3,z_2)
\otimes V({\bf Q},z_2,z_1) 
\end{array}
\label{loss11}
\end{equation}    
If we now use (\ref{Def1}), the definitions (\ref{Def2}) and (\ref{Def3}) 
of the surface correlation function and the surface excitation propagator, respectively, and the connection (\ref{impsv}) between the surface 
correlation function and the imaginary part of 
surface excitation propagator, equation (\ref{loss11}) can be written as
\begin{eqnarray}
P_{12}=-\frac{\hbar}{\pi}\sum^{\infty}_{m=-\infty}\int\frac{d{\bf Q}}{(2\pi)^2}e^{-2Qa}J^2_m({\bf Q}\brho_0)\hspace{3cm} 
\nonumber\\
\label{loss12}\\
\int\frac{d\omega}{2\pi}\ \omega_m\ [2n_1(\omega)+1]\ ImD_1({\bf Q},\omega) ImD_2({\bf Q},\omega_m).\hspace{1cm}  
\nonumber
\end{eqnarray}    
Evaluating (\ref{loss12}) we have used the fact that the real part of the function under summation and integration is 
odd and the imaginary part is an even function of $n$ and $\omega$.
$P_{12}$ is the energy transferred from the left to the right slab.
Now we have to repeat the discussion in Sec.\ref{jurniga} and substract the part of this energy which 
will be reversibly returned to the left slab. The same arguments, leading to (\ref{losse17}), will give 
this energy to be
\begin{eqnarray}
P'_{12}=\hbar\sum^{\infty}_{n=-\infty}\int\frac{d{\bf Q}}{(2\pi)^2}e^{-2Qa}J^2_n({\bf Q}\brho_0)\hspace{2cm} 
\nonumber\\
\label{loss13}\\
\int\frac{d\omega}{2\pi}\ \omega\ [2n_1(\omega)+1]\ Im D_1({\bf Q},\omega) Im D_2({\bf Q},\omega_n).  
\nonumber
\end{eqnarray}  
Expression (\ref{loss13}) represents the energy transferred from the 
left to right but which will be reversibly returned, as shown in Fig.\ref{FigA3}b. Therefore the energy 
which is irreversibly transferred from the left to the right, i.e. the dissipated power, is  
\begin{equation}
\begin{array}{c}
P_{1}=P_{12}-P'_{12}=
2\hbar\sum^{\infty}_{m=1}m\omega_0\ \int\frac{d{\bf Q}}{(2\pi)^2}e^{-2Qa}J^2_m({\bf Q}\brho_0) 
\\
\\
\int\frac{d\omega}{2\pi}\ [2n_1(\omega)+1] ImD_1({\bf Q},\omega) ImD_2({\bf Q},\omega_m).
\end{array}
\label{diss1}
\end{equation}
Analogous calculation would give the energy dissipated in the process where the charge 
fluctuation in the right slab induces fluctuations in the left slab. We have to 
exchange $1$ and $2$ in (\ref{diss1}) and replace $m\rightarrow-m$. 
Repeating the steps in (\ref{jutro}) the final result becomes:

\begin{eqnarray}
P=P_1+P_2=\hspace{3cm}
\nonumber\\
\nonumber\\
4\hbar\sum^{\infty}_{m=1}m\omega_0\ \int\frac{d{\bf Q}}{(2\pi)^2}e^{-2Qa}J^2_m({\bf Q}\brho_0)\int^{\infty}_{-\infty}\frac{d\omega}{2\pi}\hspace{2cm}
\nonumber\\
\label{njunja}\\
\left[n_1(\omega)-n_2(\omega_m)\right]ImD_1({\bf Q},\omega)ImD_2({\bf Q},\omega_m).
\nonumber
\end{eqnarray}
This expression is analogous to (\ref{jutro}). For $T=0$  $2n(\omega)+1\rightarrow sgn\omega$ and 
(\ref{njunja}) can be written as 
\begin{eqnarray}
P=\hspace{5cm}
\label{njunja1}\\
\nonumber\\
4\hbar\sum^{\infty}_{m=1}m\omega_0\ \int\frac{d{\bf Q}}{(2\pi)^2}e^{-2Qa}J^2_m({\bf Q}\brho_0)
\int^{m\omega_0}_{0}\frac{d\omega}{2\pi}\hspace{2cm}
\nonumber\\
\nonumber\\
ImD_1({\bf Q},\omega)ImD_2({\bf Q},m\omega_0-\omega).
\nonumber
\end{eqnarray}
Adding higher order terms (\ref{Eka1},\ref{Eka2})  we obtain the energy 
dissipated per unit time: 
\begin{eqnarray}
P=2\hbar\sum^{\infty}_{m=1}m\omega_0\ \int\frac{d{\bf Q}}{(2\pi)^2}e^{-2Qa}J^2_m({\bf Q}\brho_0)\times
\nonumber\\
\label{losshop}
\\
\hspace{3cm}\int^{\infty}_{-\infty}\frac{d\omega}{2\pi}C({\bf Q},\omega,\omega_m)
\nonumber
\end{eqnarray}
where $C$ is given by (\ref{kaka}). Limiting casses 
are also obtained from (\ref{losshop}). For $\omega_0=0$ and/or for $\rho_0=0$ 
obviously $P=0$.
\section{Derivation of the slab surface excitation propagators $D_{1,2}({\bf Q},\omega)$}
\label{DerofD}
\begin{figure}[h]
\centering
\includegraphics[width=4.0cm,height=7cm]{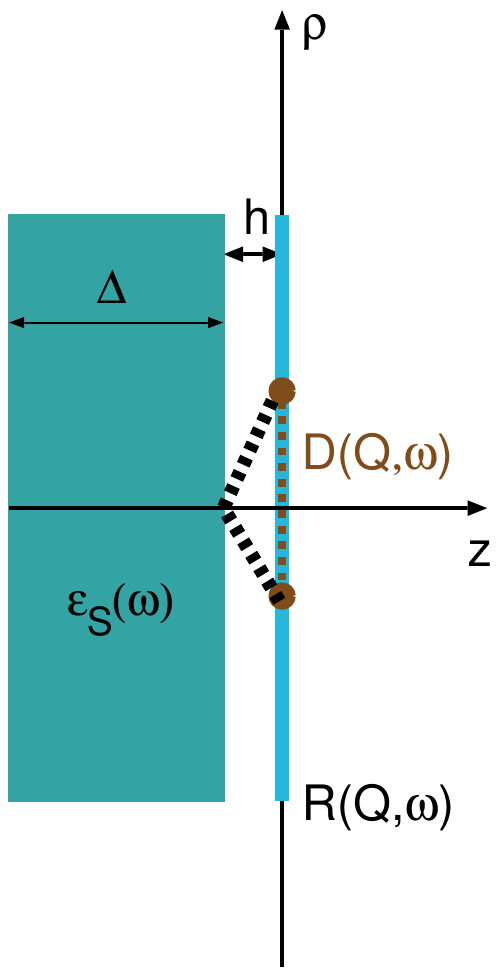}
\caption{(color online) Simplified model where the SiO$_2$ substrate is shown as a homogenous dielectric slab described by the local dielectric function $\epsilon_S(\omega)$ and graphene is described by 2D response function $R({\bf Q},\omega)$. $D({\bf Q},\omega)$ is the surface excitation propagator of the substrate/graphene composite.}
\label{Fig2}
\end{figure}

The main quantities which appear in the formula for van der Waals interaction $E_c$ or 
dissipated power $P$ are the surface excitation propagators $D_{1}({\bf Q},\omega)$ and $D_{2}({\bf Q},\omega)$ of the left (first) and right (second) slab, respectively.
The derivation of $D_1$ and $D_2$ is analogous for both slabs, so here we shall derive just one surface 
excitation propagator $D$. The structure of the monolayer-substrate composite (e.g. graphene on 
SiO$_2$) is shown in Fig.\ref{Fig2}. The slab consists of the graphene monolayer adsorbed at some small distance $h$ (e.g. $h=0.4$nm) above the substrate of macroscopic thickness $\Delta$. The 
dielectric, e.g. the SiO$_2$ slab is placed in the region $-\Delta-h\le z\le-h$ and the graphene 
layer occupies $z=0$ plane. The same model system is used in Refs.\cite{Ivan1,Ivan2} where 
the authors explore plasmon-phonon hybridization, stopping power and wake effect produced 
by the proton moving parallel to the composite. The unit cell for such huge nanostructure would 
consist of hundreds of atoms, so it is impossible to perform full {\em ab initio} ground state and structure optimization calculation. Moreover,  an {\em ab initio} calculation of the response function would be even more demanding so we need an approximation for the response function calculation. The easiest (and probably the best) approximation is to treat 
the SiO$_2$ slab as a homogeneous dielectric described by some local dielectric function $\epsilon_S(\omega)$ and to consider graphene as a purely 2D system described 
by the response function $R({\bf Q},\omega)$, as sketched in Fig.\ref{Fig2}.    
In order to derive the surface excitation 
propagator $D({\bf Q},\omega)$ we start from its definition:
\begin{eqnarray}
D({\bf Q},\omega)=v_Q\int^0_{-\infty}\ dzdz' e^{-Q(z+z')}R({\bf Q},\omega,z,z')= 
\nonumber\\
\label{defD}\\
\frac{1}{v_Q}\left\{W({\bf Q},\omega,z=0,z'=0)-v_Q\right\};\  i=1,2.
\nonumber
\end{eqnarray}
which connects the surface excitation propagator with the screened Coulomb interaction $W({\bf Q},\omega,z=0,z'=0)$ at $z=z'=0$ surface.
Here $R({\bf Q},\omega,z,z')$ represents the nonlocal dielectric function of graphene/dielectric composite which we assume occupies the region $z,z'\le 0$. 

It is well known  \cite{gr2D1,gr2D2,gr2D3,gr2D4} that physical properties of a graphene 
monolayer in the low ($Q,\omega$) region can be described to a very good approximation 
assuming the monolayer to be strictly twodimensional, so that the nonlocal independent 
electron response function can be written as
\begin{equation}
R^{0}({\bf Q},\omega,z,z')=R^{0}({\bf Q},\omega)\delta(z)\delta(z') 
\end{equation}
where we assume that the graphene lies in the $z=0$ plane and the response function 
$R^{0}({\bf Q},\omega)$ can be derived from first 
principles, as decribed in Sec.\ref{nabnak}. 
Dynamically screened response function $R({\bf Q},\omega)$ in RPA is given as 
a series of terms
\begin{equation}
R({\bf Q},\omega)=R^{0}+R^{0}v_QR^{0}+...=\frac{R^0({\bf Q},\omega)}{1-v_QR^0({\bf Q},\omega)}.
\label{RPARR}
\end{equation} 
If we assume for the moment that there is no dielectric in the system (e.g. $\epsilon_S(\omega)=1$) then the screened Coulomb 
interaction is simply given by 
\begin{equation}
W({\bf Q},\omega,z=0,z'=0)=v_Q+v_QR({\bf Q},\omega)v_Q.
\label{scrCint}
\end{equation}
Using the definition (\ref{defD}) the surface excitation propagator becomes
\begin{equation}
D({\bf Q},\omega)=v_QR({\bf Q},\omega).
\end{equation}
When the dielectric slab is introduced, the external charges and charge density fluctuations in the graphene layer do not interact via the bare Coulomb 
interaction $v_Q$ but via the Columob interaction modified by the presence of 
the dielectric slab \cite{Laplace}     
\begin{equation}
v_Q\rightarrow\tilde{v}_Q(\omega)=v_Q\left[1+D_S({\bf Q},\omega)\right],
\label{tildeV11}
\end{equation}
where the substrate surface excitation propagator is
\begin{equation}
D_S({\bf Q},\omega)=D_S(\omega)\frac{1-e^{-2Q\Delta}}{1-D^2_S(\omega)e^{-2Q\Delta}}e^{-2Qh}
\label{labina}
\end{equation}
and
\begin{equation}
D_S(\omega)=\frac{1-\epsilon_S(\omega)}{1+\epsilon_S(\omega)}
\end{equation}   
represents the surface excitation propagator of a semiinfinite 
($\Delta\rightarrow\infty$, $h=0$) dielectric.
This causes that the screened Colulomb interaction (\ref{scrCint}) becomes the function 
of $\tilde{v}_Q(\omega)$
\begin{equation}
W\rightarrow\tilde{W}=\tilde{v}_Q(\omega)+\tilde{v}_Q(\omega)\tilde{R}({\bf Q},\omega)\tilde{v}_Q(\omega),
\label{modifiedW}
\end{equation}
where, because charge density fluctuations inside graphene also interact via $\tilde{v}_Q(\omega)$, the screened response function is modified as
\begin{equation}
\tilde{R}({\bf Q},\omega)=\frac{R^0({\bf Q},\omega)}{1-\tilde{v}_Q(\omega)R^0({\bf Q},\omega)}.
\label{curoni}
\end{equation}
Finally, after inserting (\ref{modifiedW}) into (\ref{defD}) we obtain the surface excitation propagator in the presence of 
the dielectric
\begin{eqnarray}
D({\bf Q},\omega)=\frac{1}{v_Q}\left\{\tilde{v}_Q(\omega)\tilde{R}_i({\bf Q},\omega)\tilde{v}_Q(\omega)+\right.
\label{newD}\\
\hspace{3cm}\left.\tilde{v}_Q(\omega)-v_Q\right\}.
\nonumber
\end{eqnarray}
which can be rewritten in a more transparent form as
\begin{eqnarray}
D({\bf Q},\omega)=\hspace{5cm}
\nonumber\\
\nonumber\\
\frac{D_S({\bf Q},\omega)+v_QR({\bf Q},\omega)+2v_QR({\bf Q},\omega)D_S({\bf Q},\omega)}{1-v_QR({\bf Q},\omega)D_S({\bf Q},\omega)}.
\end{eqnarray}
The spectrum of coupled excitations in a single slab can be calculated from
\begin{equation}
S({\bf Q},\omega)=-\frac{1}{\pi}ImD({\bf Q},\omega).
\end{equation}
For the coupled slabs described by their surface excitations propagators 
$D_1$ and $D_2$, separated by the distance $a$, in a similar way we can derive 
the propagator $\tilde{D}$ for the coupled system 
\begin{equation}
\tilde{D}({\bf Q},\omega)=
\frac{D_1({\bf Q},\omega)+D_2({\bf Q},\omega)+2D_1({\bf Q},\omega)D_2({\bf Q},\omega)}{1-e^{-2Qa}D_1({\bf Q},\omega)D_2({\bf Q},\omega)}
\end{equation}
and the excitation spectrum of this system is
\begin{equation}
\tilde{S}({\bf Q},\omega)=-\frac{1}{\pi}Im\tilde{D}({\bf Q},\omega).
\end{equation}
\section{Description of substrate and graphene dynamical response}
\label{nabnak}
The results in Sec.\ref{DerofD} are quite general and can be applied to a monolayer 
of any material on any dielectric substrate.
Now we shall specify the dielectric substrate to be the homogenous 
layer of ionic crystal SiO$_2$.

Dielectric properties (or dynamical response) of bulk ionic crystals in the long-wavelength limit can 
be described in terms of their optical phonons at the $\Gamma$ point. More complex polar crystals such as SiO$_2$ possess a multitude of different optical 
phonons of different symmetries and polarizations. However, here we suppose  that 
SiO$_2$ posses  two well-defined, non-dispersing transverse
optical (TO) phonon modes at the frequencies $\omega_{TO1}$
and $\omega_{TO2}$ with the corresponding damping rates
$\gamma_{TO1}$ and $\gamma_{TO2}$, giving rise to a 
dielectric function of the form \cite{Ivan1,Ivan2}
\begin{eqnarray}
\epsilon_S(\omega)=\epsilon_{\infty}+(\epsilon_i-\epsilon_{\infty})\frac{\omega^2_{TO2}}{\omega^2_{TO2}-\omega^2-i\omega\gamma_{TO2}}+
\nonumber\\
(\epsilon_0-\epsilon_i)\frac{\omega^2_{TO1}}{\omega^2_{TO1}-\omega^2-i\omega\gamma_{TO1}} ,
\label{dielectric}
\end{eqnarray}
where $\epsilon_0$, $\epsilon_i$, and $\epsilon_{\infty}$ represent  
the dielectric constant for SiO$_2$ at the zero, intermediate, and very large
frequencies. This dielectric function will be inserted in the expression (\ref{labina}) for the substrate surface excitation propagator $D_S({\bf Q},\omega)$. 

The graphene response function $R({\bf Q},\omega)$ is given by (\ref{curoni}) in terms of the noninteracting response function
\begin{equation}
R^{0}({\bf Q},\omega)=L\ R^{0}_{{\bf G}=0{\bf G}'=0}({\bf Q},\omega)
\label{Chi02D}
\end{equation}
where the 3D Fourier transform of independent electron response function is given 
by \cite{PRB13} 
\begin{eqnarray}
R^{0}_{{\bf G}{\bf G}'}({\bf Q},\omega)=\hspace{5cm}
\nonumber\\
\frac{2}{\Omega}\sum_{{\bf K}\in S.B.Z.}\sum_{n,m}\ \frac{f_n({\bf K})-f_m({\bf K}+{\bf Q})}
{\hbar\omega+i\eta+E_n({\bf K})-E_m({\bf K}+{\bf Q})}\times
\label{Resfun0}\\
\rho_{n{\bf K},m{\bf K}+{\bf Q}}({\bf G})\ \rho^*_{n{\bf K},m{\bf K}+{\bf Q}}({\bf G'}),
\nonumber
\end{eqnarray}  
where $f_{n{\bf K}}=[e^{(E_{n{\bf K}}-E_F)/kT}+1]^{-1}$ is the Fermi-Dirac distribution at 
temperature $T$. The charge vertices in (\ref{Resfun0}) have the form 
\begin{equation}
\rho_{n{\bf K},m{\bf K}+{\bf Q}}({\bf G})=
\int_\Omega\ d{\bf r}e^{-i({\bf Q}+{\bf G}){\bf r}}\ \phi^*_{n{\bf K}}({\bf r})\phi_{n{\bf K}+{\bf Q}}({\bf r})
\label{Matrel}
\end{equation}
where ${\bf Q}$ is the momentum transfer vector parallel to the $x-y$ plane, ${\bf G}=({\bf G}_\parallel,G_z)$ are $3D$ reciprocal lattice vectors and 
${\bf r}=(\brho,z)$ is a $3D$ position vector. Integration in (\ref{Matrel}) is performed over the normalization volume $\Omega=S\times L$, where $S$ is the 
normalization surface and $L$ is the superlattice constant in $z$ direction (separation between graphene layers is superlattice arrangement). 
Plane wave expansion of the wave function has the form 
\[
\phi_{n{\bf K}}(\brho,z)=\frac{1}{\sqrt{\Omega}}e^{i{\bf K}\brho}\ \sum_{\bf G}C_{n{\bf K}}({\bf G})e^{i{\bf G}{\bf r}},
\]
where the coefficients $C_{n{\bf K}}$ are obtained by solving the Local Density Approximation-Kohn Sham (LDA-KS) equations selfconsistently as will be discussed below.
However, this straightforward calculation of graphene response functions $R({\bf Q},\omega)$ is not 
sufficient if we want to investigate the hybridization between the Dirac plasmon and Fuchs-Kliewer (FK) phonons at dielectric surfaces. Namely, due to 
the very low energy of FK phonons ($\sim 50$meV) the crossing of their dispersion relations with Dirac plasmon occurs for very small wave vectors ($Q<0.001$a.u.). 
On the other hand even for very dense $K$-point mesh sampling, as for example $601\times 601\times 1$ used in this calculation, 
the minimum transfer wave vector $Q$ which can be reached (e.g. $Q=0.0026$a.u.$^{-1}$ in this calculation) is still bigger than FK phonon-Dirac 
plasmon crossing wave vector. Therefore we have to find the way how to calculate $R({\bf Q},\omega)$ for a denser Q-point mesh 
in the optical $Q\approx 0$ limit. One possible way is that instead of calculating response function $R^{0}({\bf Q},\omega)$ we calculate the optical 
($Q=0$) conductivity $\sigma(\omega)$. The optical conductivity in graphene can be written as \cite{gr2D3}    
\begin{equation}
\sigma(\omega)=\sigma^{\mathrm{intra}}(\omega)+\sigma^{\mathrm{inter}}(\omega),       
\label{curren1}
\end{equation}
where 
\begin{equation}
\sigma^{\mathrm{intra}}(\omega)=\frac{i\rho_0}{\omega+i\eta_{\mathrm{intra}}}
\label{curren2}
\end{equation}
is intraband or Drude conductivity and where 
\begin{equation}
\rho_0=-\frac{2}{\Omega}\sum_{{\bf K},n}\frac{\partial f^i_n({\bf K})}{\partial E_n({\bf K})}
|j^{x}_{n{\bf K},n{\bf K}}({\bf G}=0)|^2
\label{curren3}
\end{equation}
represents the effective number of charge carriers. 
The interband conductivity is  
\begin{eqnarray}
\sigma^{\mathrm{inter}}(\omega)=
\frac{-2i}{\omega\Omega}\sum_{{\bf K},n\neq m}\ \frac{\hbar\omega}{E_n({\bf K})-E_m({\bf K})}\times
\nonumber\\
\frac{f^i_n({\bf K})-f^i_m({\bf K})}
{\hbar\omega+i\eta_{\mathrm{inter}}+E_n({\bf K})-E_m({\bf K})}\times
\label{curren4}\\
j^{x}_{n{\bf K},m{\bf K}}({\bf G}=0)\ [j^{x}_{n{\bf K},m{\bf K}}({\bf G}'=0)]^*
\nonumber
\end{eqnarray}
where the current vertices are given by
\begin{equation}
j^{\mu}_{n{\bf K},m{\bf K}+{\bf Q}}({\bf G})=
\int_\Omega\ d{\bf r}e^{-i({\bf Q}+{\bf G}){\bf r}}\ 
j^{\mu}_{n{\bf K},m{\bf K}+{\bf Q}}({\bf r}),
\label{curren5}
\end{equation}
and 
\begin{eqnarray}
j^{\mu}_{n{\bf K},m{\bf K}+{\bf Q}}({\bf r})=
\frac{\hbar e}{2im}
\left\{\phi_{n{\bf K}}^*({\bf r})\partial_\mu\phi_{m{\bf K}+{\bf Q}}({\bf r})\right.\hspace{2cm}
\label{curren6}\\
\hspace{2cm}-\left.[\partial_\mu\phi_{n{\bf K}}^*({\bf r})]\phi_{m{\bf K}+{\bf Q}}({\bf r})\right\}.
\nonumber
\end{eqnarray}
In the optical $Q\approx 0$ limit the  independent electron response function can be written in terms of optical 
conductivities (\ref{curren1}) as \cite{Zoran}
\begin{equation}
R^0({\bf Q}\approx 0,\omega)=L\ \frac{Q^2}{i\omega}\sigma(\omega).
\label{chi-sigma}
\end{equation}
Finally, the RPA or screened response function $R({\bf Q},\omega)$ can be obtained from (\ref{chi-sigma}) 
using (\ref{RPARR}). 

In the calculation of Sec.\ref{resuuult} we shall assume the 
graphene response to be isotropic in the small $({\bf Q},\omega)$ limit. This means 
that the graphene response functions and the corresponding surface excitation functions are functions of 
$Q$ and not of ${\bf Q}$.      

\subsection{Computational details}
\label{Comp}
The first part of the calculation consists of determining the KS ground state of the single layer graphene and the corresponding wave functions 
$\phi_{n{\bf K}}(\brho,z)$ and energies $E_n({\bf K})$. For graphene unit cell constant we use the experimental value of $a=4.651\ \mathrm{a.u.}$ \cite{lattice}, and superlattice unit cell constant (separation of graphene layers) is $L=5a$. For calculating KS wave functions and energies we use a plane-wave self-consistent field DFT code (PWSCF) within the QUANTUM ESPRESSO (QE) package \cite{QE}. The core-electron interaction was approximated by the norm-conserving pseudopotentials \cite{normcon}, and the exchange correlation (XC) potential by the Perdew-Zunger local density approximation (LDA) \cite{lda1}. To calculate the ground state electronic density we use $21\times21\times1$ Monkhorst-Pack K-point 
mesh \cite{MPmesh} of the first Brillouin zone (BZ) and for the plane-wave cut-off energy we choose 50 Ry. 
The second part of calculation consists of determining the independent electron response function (\ref{Resfun0}) and conductivity 
(\ref{curren1}--\ref{curren4}). In order to achieve better resolution in the long wavelength ($Q\approx 0$) and low energy ($\omega\approx 0$) 
limit the response function (\ref{Resfun0},\ref{Matrel}) and conductivity (\ref{curren1}--\ref{curren6}) are evaluated from the wave functions $\phi_{n{\bf K}}({\bf r})$ and energies $E_n({\bf K})$ calculated for the $601\times601\times1$ Monkhorst-Pack K-point mesh  which coresponds to 361801 K-points in the first Brillouin zone (1BZ). Band summations ($n,m$) in (\ref{Resfun0}), (\ref{curren3}) and (\ref{curren4}) are performed over 30 bands. In the calculation we use two kinds of damping parameters: $\eta_{\mathrm{intra}}=10$meV for transitions within the same bands ($n\leftrightarrow n$), and $\eta_{\mathrm{inter}}=50$meV for transitions between different bands ($n\leftrightarrow m$). 
For bulk SiO$_2$ dielectric function given by (\ref{dielectric}) we use the following 
parameters: $\epsilon_0=3.9$, $\epsilon_i=3.05$, $\epsilon_{\infty}=2.5$, 
$\omega_{TO1}=55.6$ meV, $\omega_{TO2}=138.1$ meV, $\gamma_{TO1}=5.368$ meV and $\gamma_{TO2}=8.947$ meV taken from Ref.\cite{Dielectricpar}. 
For the gap between graphene and the SiO$_2$ surface, we take $h=4\AA[7.55$ a.u.$]$ \cite{height}.

\section{Results for graphene monolayers on SiO$_2$ substrates}
\label{resuuult}
Theoretical expressions derived in Sec.\ref{Sec2} (and in Appendix \ref{AppA}) are quite 
general, i.e. are valid for any pair of crystal slabs described by their response 
functions, while the corresponding surface excitation functions derived in 
Sec.\ref{DerofD} are valid for any 2D adsorbed monolayer on any dielectric 
substrate. In this section we shall apply these results to calculate reactive and 
dissipative response of various combinations of slabs consisting of graphene monolayers with variable doping on SiO$_2$ substrate, using the dynamical surface response functions of 
these materials given in Sec.\ref{DerofD}.          

\begin{figure}[t]
\includegraphics[width=6cm,height=5cm]{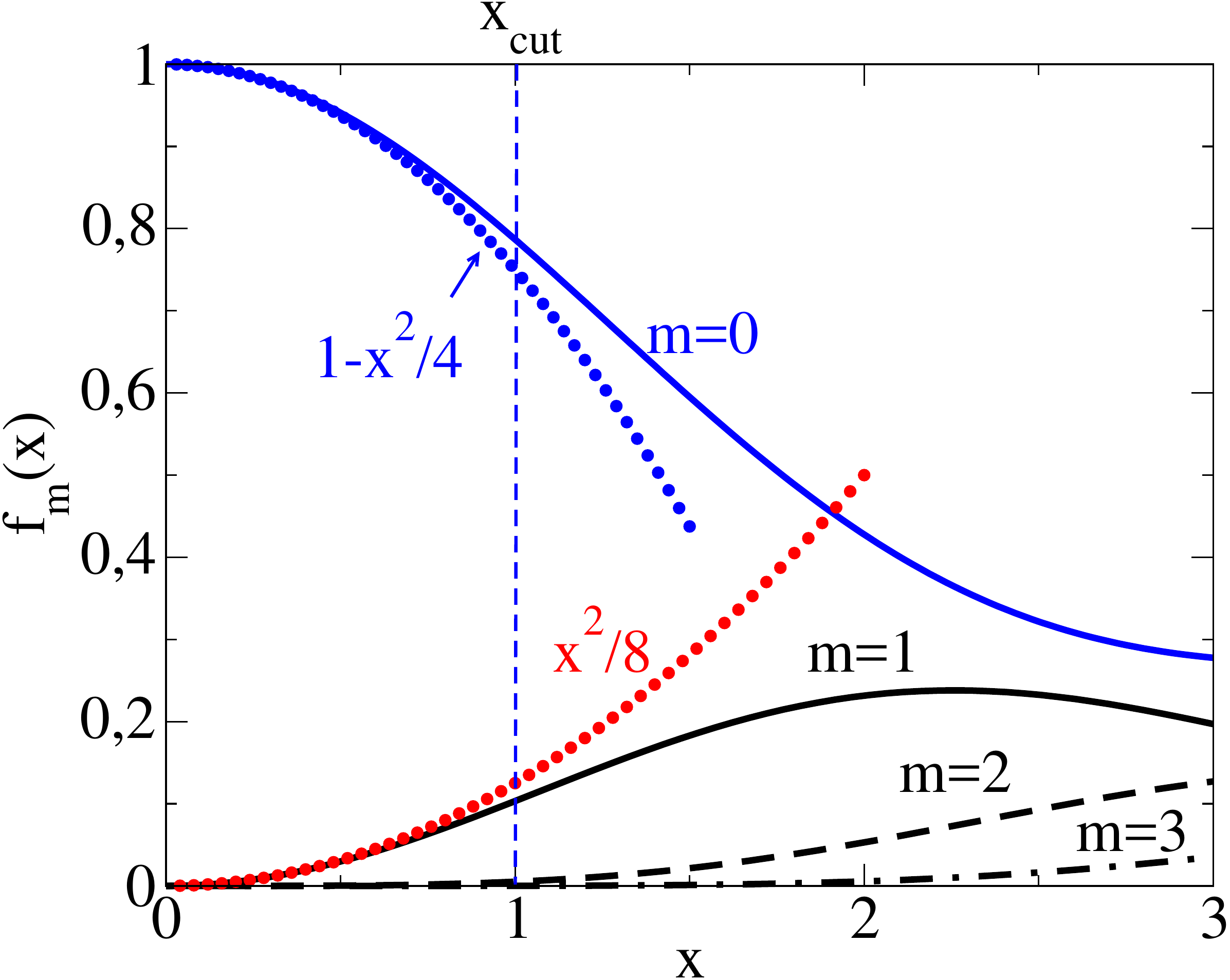}
\caption{Function $f_m(x)$ for  m=0 (blue solid line), $m=1$ (black solid line), $m=2$ (black dashed line) and $m=3$ (black dashed-dotted line). Vertical dashed line denotes the maximum argument $x_{cut}$ defined by parameters ($a$ and $\rho_0$) used in the calculation.}
\label{Fig3}
\end{figure}

Before proceeding with detailed calculations a few general comments are 
in order. Though the derived expressions for van der Waals and dissipated 
power (\ref{vdwFIN}) and (\ref{losshop}), respectively, include temperature 
dependence, in the systems studied here inclusion of finite temperature leads to practically 
no effects, therefore all results will be reported for $T=0$. 
The dependence of these two physical properties on the two parameters, the distance 
between the slabs $a$ and the oscillation amplitude $\rho_0$, can be analyzed if 
we recognize in the expressions  (\ref{vdwFIN}) and (\ref{losshop}) the function     
\begin{equation}
f_m(x)=\int^{2\pi}_0\frac{d\phi}{2\pi}J^2_m(x\cos\phi),
\label{fm}
\end{equation}
which is possible because of the assumed isotropy of graphene response. The function 
$f_m(x)$ is shown in Fig.\ref{Fig3} for first four $m$'s, where $x=Q\rho_0$. 
Another important  factor in (\ref{vdwFIN}) and (\ref{losshop}) is $e^{-2Qa}$  which defines the 
cutoff wave vector $Q_c$, depending on the slab separation $a$. 
The separations we shall consider in this calculation are $a=10-50$nm which defines the cutoff wave 
vector $Q_c\approx 0.05a.u.$. On the other hand, the ampitudes which will be considered are 
$\rho_0\approx 0.1-1$nm. This finally provides the maximum argument $Q$ of the functions (\ref{fm}) which is $x_{cut}\approx 1$. From Fig.\ref{Fig3} is obvious that up to $x_{cut}$ only the $m=0$ 
and $m=1$ terms will contribute. Moreover, for $x<x_{cut}$ the Bessels functions can be approximated 
as $J_0\approx 1-\frac{x^2}{4}$ and $J_m(x)\approx \frac{x^m}{2^mm!};\ m>1$ and therefore  
\begin{equation}
f_0\approx 1-\frac{x^2}{4};\ \ \ f_1(x)\approx \frac{x^2}{8}. 
\label{approx}
\end{equation}
In Fig.\ref{Fig3} we see that approximation (\ref{approx}) is valid almost up 
to $x_{cut}$.

\subsection{Spectra of coupled modes}
\label{cmspe}
 
In this section we shall first discuss the spectra of coupled plasmon/phonon excitations 
in one and two graphene/SiO$_2$ slabs separated by distance $a$ in order to understand 
the dominant dissipation mechanisms.

\begin{figure*}[tt]
\includegraphics[width=1.0\columnwidth]{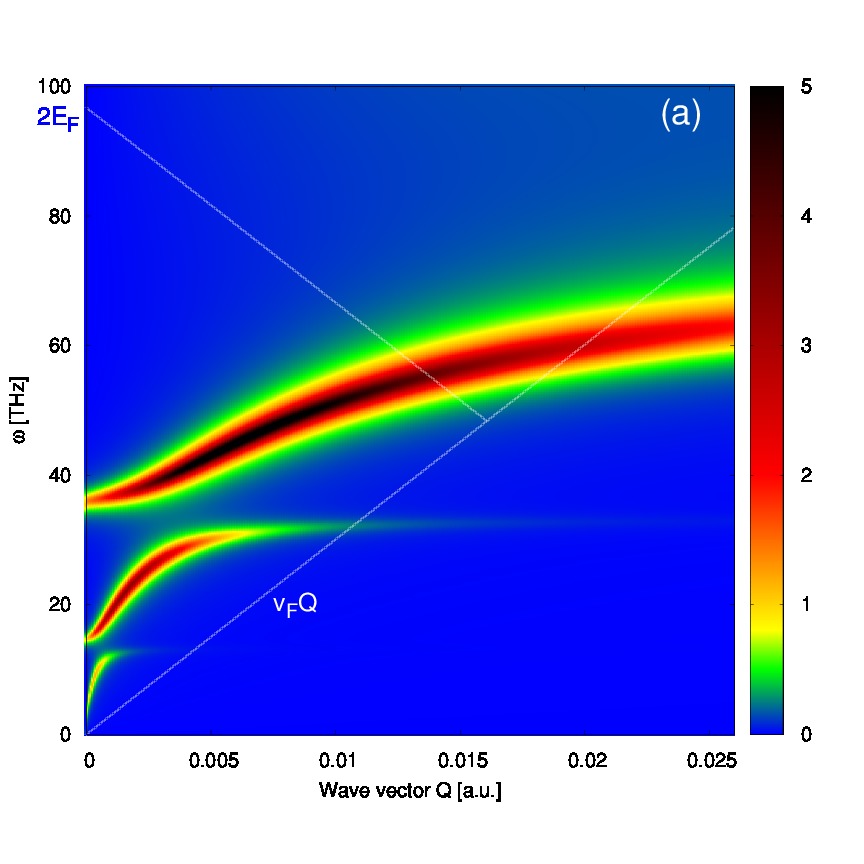}
\includegraphics[width=1.0\columnwidth]{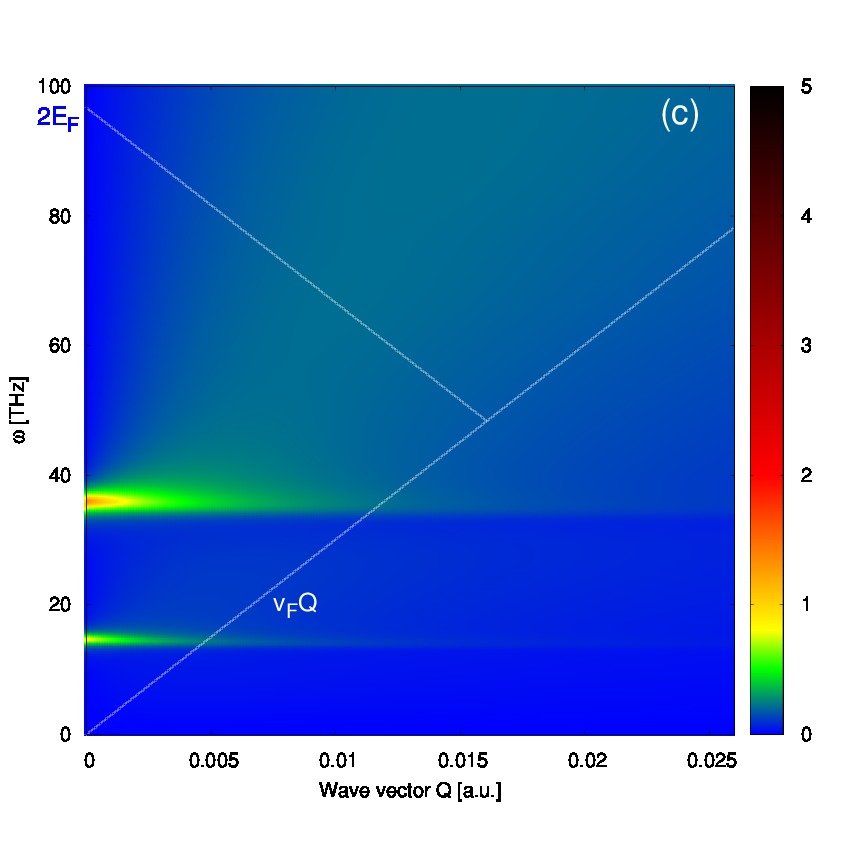}
\includegraphics[width=1.0\columnwidth]{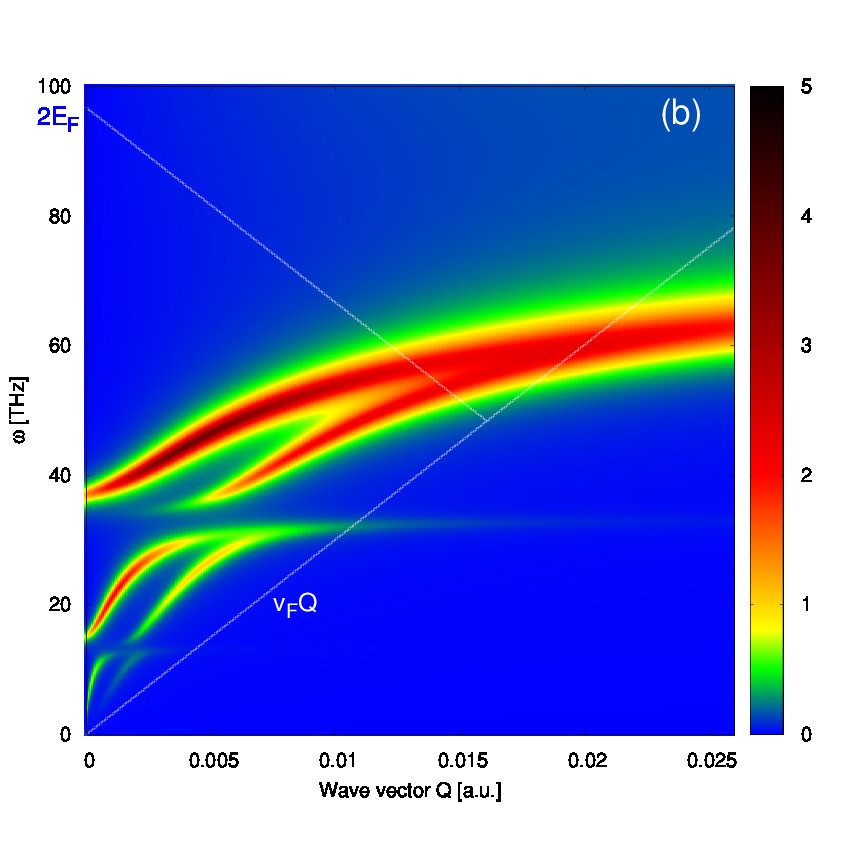}
\includegraphics[width=1.0\columnwidth]{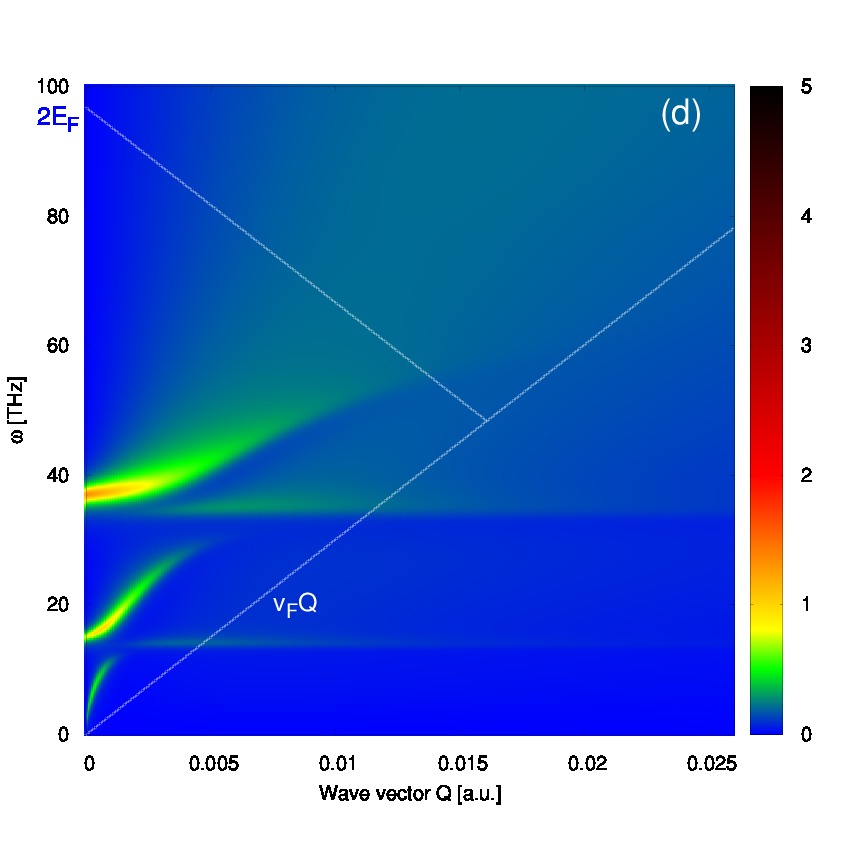}
\caption{(Color online) The spectra of surface excitations in (a) graphene(200meV)/SiO$_2$ single 
slab (as shown in Fig.\ref{Fig2}), (b) in the system consisting of two equal graphene(200meV)/SiO$_2$ slabs, (as shown in Fig.\ref{Fig1}) separated by distance 5nm, (c) single 
graphene(0meV)/SiO$_2$ slab and (d) in the system consisting of two unequal slabs, 
graphene(200meV)/SiO$_2$ and graphene(0meV)/SiO$_2$, separated by distance 5nm.} 
\label{Fig4}
\end{figure*}

Fig.\ref{Fig4}(a) shows the spectrum of surface  excitations $S(Q,\omega)=-Im D(Q,\omega)$ 
in graphene(200meV)/SiO$_2$ slab (as shown in Fig.\ref{Fig2}) and Fig.\ref{Fig4}(b) in the system 
which consists of two graphene/SiO$_2$ slabs (as shown in Fig.\ref{Fig1}) separated by distance $a=5$nm. In the lonwavelength limit the SiO$_2$ surface suports two surface polar (FK) TO phonons with 
flat dispersions and the doped graphene contains a Dirac plasmon with square root dispersion. 
Coupling between these modes results in three branches, as shown in Fig.\ref{Fig4}(a). For larger 
$Q$ the first and second flat branches are phononlike, i.e. their induced 
electrical fields mostly come from polarization modes on the dielectric
surface. On the other hand, the third square root branch is 
plasmon-like, i.e. its induced electrical field mostly comes from charge density 
oscillations localised in the graphene layer. However, in the $Q\rightarrow 0$ limit 
the strong hybridization (avoided crossings) between these modes occur and they 
possess mixed plasmon-phonon character. When another slab is brought in the vicinity the 
three modes in each slab interact which results in the mode splitting and formation of 
six coupled modes as shown in Fig.\ref{Fig4}(b).    
Figure \ref{Fig4}(c) shows the spectrum of surface  excitations in the 
graphene($0$meV)/SiO$_2$ slab. Because the pristine graphene does not support Dirac plasmon 
the spectrum consist just of two weak phonon branches $\omega_{TO1}$ and $\omega_{TO2}$ damped 
by $\pi\rightarrow\pi^*$ excitations. The spectrum of surface  excitations in two equal 
graphene(0meV)/SiO$_2$ slabs separated by 5nm (not shown here) is very similar to the one 
shown in Fig.\ref{Fig4}(c) which indicates weak interaction between phonons 
in the two slabs. This could be the consequence of strong screening of FK phonons 
by graphene adlayers which reduces the range of their induced electrical 
field. Figure \ref{Fig4}(d) shows the spectrum in the system which consists of two different 
slabs, graphene($0$meV)/SiO$_2$ and graphene($200$meV)/SiO$_2$, separated by $5$nm. 
One can notice interesting hybridization between the Dirac plasmon and two phonons in one slab and two phonons in another slab giving five branches. 
In the next section we shall explore how particular plasmon-phonon modes contribute to the 
dissipated power in two oscillating slabs. 

\subsection{Modification of van der Waals force}
\label{weakvdW}
Van der Waals energy and attractive force are usually calculated and measured for static 
objects. Here we show how their relative oscillating motion can reduce this attraction, 
which can be relevant not only from the theoretical standpoint but 
also in some experimental situations and applications. This phenomenon is present 
also in the case of parallel motion, as shown in the Appendix \ref{AppA}, but 
this situation would be more difficult to realize in practice. 

Making use of the approximation (\ref{approx}) for the lowest order 
terms of the functions $f_0$ and $f_1$ given by (\ref{fm}) we can 
rewrite the expression (\ref{vdwFIN}) for the van der Waals energy as  
\begin{eqnarray}
E_{c}(a)=\frac{\hbar}{2}\int\frac{QdQ}{2\pi}
\int^{\infty}_{-\infty}\frac{d\omega}{2\pi}\hspace{3cm}
\label{java}\\
\nonumber\\
\left\{ \left[1-\frac{Q^2\rho_0^2}{4}\right]A(Q,\omega,\omega)+\frac{1}{4}Q^2\rho_0^2\ A(Q,\omega,\omega-\omega_0)\right\}
\nonumber
\end{eqnarray}
where $A$ is given by (\ref{jalko}) and (\ref{jujucka}).  
In the $T\rightarrow 0$ limit and neglecting higher order terms $A$ reduces to
\begin{eqnarray}
A(Q,\omega,\omega')=\hspace{5cm}
\nonumber\\
e^{-2Qa}sgn\omega\left\{ImD_1(Q,\omega)ReD_2(Q,\omega')+(1\leftrightarrow 2)\right\}
\nonumber
\end{eqnarray}
We see that for $\rho_0\rightarrow 0$ the van der Waals energy reduces 
to the standard result for the static case, and for $\rho_0\ne 0$ and 
$\omega_0\ne 0$ the lowest order corrections scale with $\rho^2_0$.    
From (\ref{java}), and also from (\ref{njunja2}), we see that the slab separation $a$ (because of exponential 
factor $e^{-2Qa}$) reduces the wave vector range to $Q<1/2a$.

\begin{figure}[t]
\includegraphics[width=1.0\columnwidth]{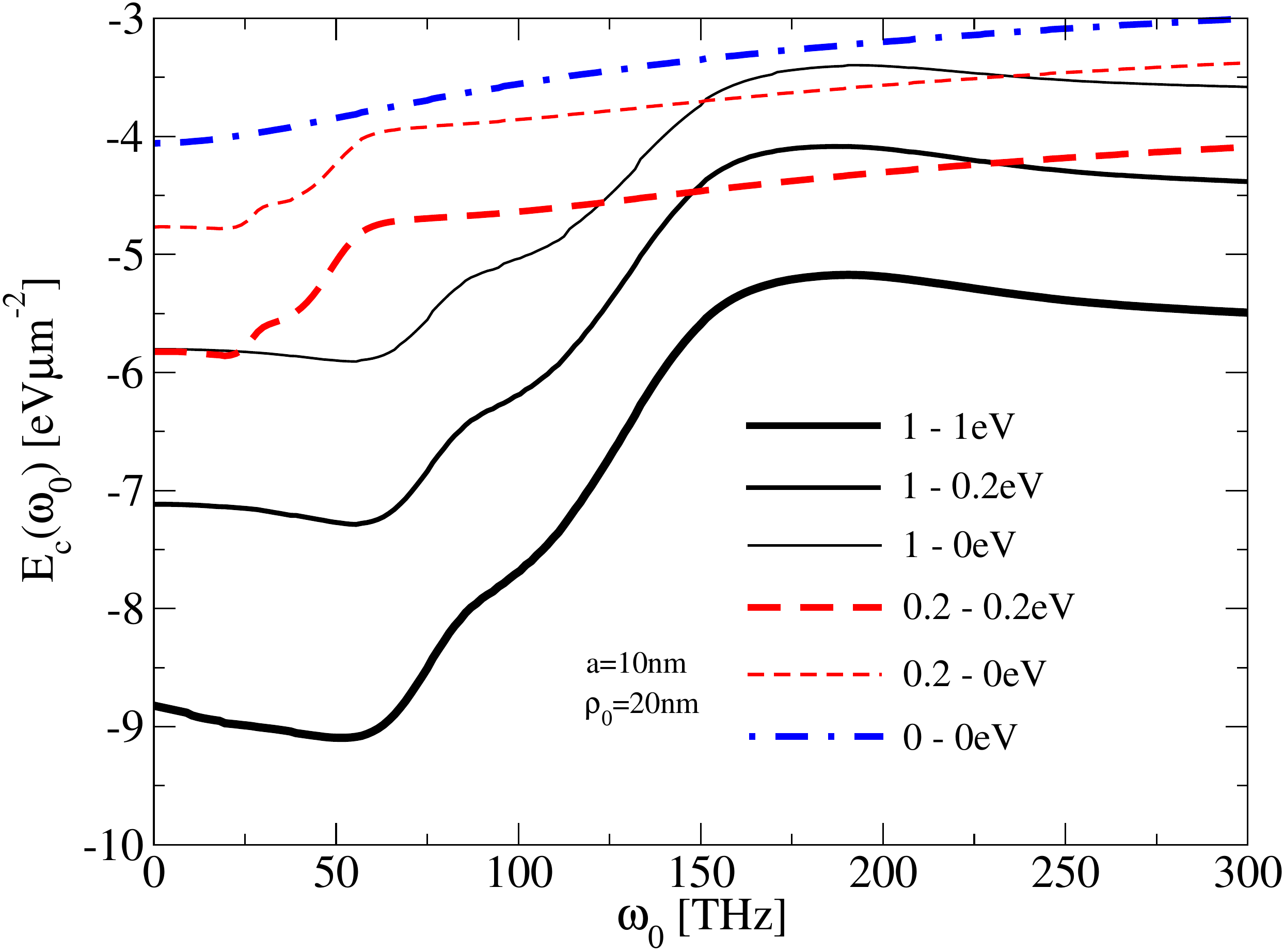}
\caption{(Color online). Van der Waals energies $E_c$ of two variously doped, unsupported full conductivity  
(\ref{curren1}--\ref{curren4}) graphenes as functions of driving frequency $\omega_0$.
The left-right graphene dopings are $1-1$eV (thick black solid),  $1-0.2$eV (black solid), 1-1eV (thin black 
solid), $0.2-0.2$eV (red dashed), $0.2-0$eV (thin red dashed), $0-0$eV (blue dashed-dotted), as also denoted in the figure.
Separation between graphenes is $a=10$nm and oscillating amplitude is $\rho_0=20$nm. } 
\label{Fig5}
\end{figure}

Fig.\ref{Fig5} shows van der Waals energies $E_c$ of two variously doped, unsupported full conductivity 
(\ref{curren1}--\ref{curren4}) graphenes as functions of the driving frequency $\omega_0$. 
The driving amplitude is $\rho_0=20$nm and separation between slabs is $a=10$nm. 
For the case of two heavily and equally doped graphenes $1-1$eV (thick black solid line)
the 'static' ($\omega_0=0$) van der Waals energy is the largest in comparison with other doping combinations. This is reasonable considering that 
then except of $\pi$ and $\pi+\sigma$ plasmons (and corresponding electron-hole 
excitations) the graphenes support strong Dirac plasmons which are all in resonance. 
Therefore, the charge density fluctuation in one slab $ImD_1(\omega)$ resonantly induces electrical field in another slab 
$ReD_2(\omega)$ to which it couples, and 
vice versa.  
As the driving frequency $\omega_0$ increases the  fluctuation and the induced field do not match any more, i.e. $ImD_1(\omega)$ and $ReD_2(\omega+n\omega_0)$ become 
Doppler shifted and vdW energy is expected to decrease. 
However, the vdW energy first exhibits a wide plateau until $\omega_0<50$THz. 
We performed a  separate vdW energy calculation for two unsupported 
Drude (\ref{curren1},\ref{curren2}) graphenes (not shown here) and noticed that it shows the same 
features as presented in Fig.\ref{Fig5}. This suggests that Dirac plasmons are responsible for all characteristic 
features in vdW energy (for larger dopings).
Therefore, the plateau arises probably because the Dirac plasmon fluctuation in one slab, e.g. at $\omega_p$, can be efficiently screened by induced 
plasmon field in another slab which is not necessarily at the same frequency $\omega_p$. 
Moreover, graphene, regardless of doping, exhibits perfect screening $ReD(Q\approx 0,\omega\approx 0)\approx -1$ \cite{Duncan2012} causing that 
the static point charge feels image potential. This causes that $E_c$ shows almost identical plateau 
for the case of differently doped graphenes $1-0.2$eV (black solid line) and $1-0eV$ 
(thin black solid line). 
As the doping difference increases plateau energy decreases which is 
reasonable because of plasmon resonance breakdown. 
For larger $\omega_0>50$THz  the Dirac plasmon in one slab does not match any more the 
perfect screening regime in another one, resulting in a rapid decrease or weakening of vdW energy.
In the case of weakly doped graphenes, such as the combinations $0.2-0.2$eV (red dashed line) and $0.2-0$eV  (thin red dashed 
lines), the 'static' $\omega_0\approx 0$  van der Waals energy reduces in comparison with the heavy
doping (combinations with $1$eV) cases. This is reasonable considering that Dirac plasmon spectral weight decreases with 
doping. Additionally, it can be noted that for lower doping the vdW plateau shifts to $\omega_0<25$THz. 
This is because the perfect screening frequency region can be roughly estimated as 
$ReD(\omega<\omega_p)\approx-1$, so, as the plasmon energy decreases the frequency interval whithin which fluctuations are perfectly screened becomes narrower.   
It is interesting to notice that for some frequencies (e.g. $\omega_0>100$THz) the resonant but low doping vdW energy (e.g.  $0.2-0.2$eV case) 
overcomes the heavily doped but off resonance vdW energy (such as the cases $1-0.2$eV and $1-0$eV). 
The static $\omega_0=0$ vdW energy of pristine graphenes $0-0$eV (blue dashed dotted line) is the weakest and shows smooth decreasing, almost 
linear behaviour. In this case there  are no Dirac plasmons in the graphenes spectra. Therefore,  only resonant coupling between $\pi\rightarrow\pi^*$ electron-hole 
excitations, $\pi$ and $\pi+\sigma$ plasmons contribute to the vdW energy. As the frequency $\omega_0$ increases the overlap between these 
electronic excitations decreases causing smooth and linear vdW energy weakening.   
The same linear behaviour (for $\omega_0>50$THz) can be noticed for doping combinations $0.2-0.2$eV and  $0.2-0$eV  
which proves that for lower dopings the dominant vdW energy weakening mechanism  becomes off-resonant coupling 
between $\pi\rightarrow\pi^*$ electron-hole excitations, $\pi$ and $\pi+\sigma$ plasmons. 

It should be noted here that such designed (graphene based) slabs might enable modification of attraction between slabs,  e.g. controlled 'sticking' 
and 'un-sticking' of two slabs. For example, two heavily doped graphenes ($1-1$eV case in Fig.\ref{Fig5}) are strongly bound, however 
binding energy between pristine graphenes ($0-0$eV case achieved, e.g. simply by electrostatic gating) is reduced more than 
twice. Moreover, for larger $\omega_0$ (and fixed doping) the dynamical binding energy is substantially reduced, leading to 'un-sticking' 
of two slabs, and vice versa, their 're-sticking' by reducing the driving frequency.

\subsection{Dissipated power - substrate dependence}
\label{DISsub}
In this section we shall explore how the dissipation power in two oscillating 
slabs depends on the conductivity model we use to describe graphene and how 
substrate influences the dissipation power.

In order to facilitate the analysis of the results we shall again use the 
approximation (\ref{approx}). The lowest order term 
which contributes in (\ref{losshop}) is $f_1$, and from Fig.\ref{Fig3} it is 
obvious that, for $x<x_{cut}$, the higher order terms ($m=2,3,...$) do not 
contribute and $f_1$ can be freely approximated by (\ref{approx}) (red dotted line).
Furthermore, because the higher order processes (see Fig.\ref{FigA2}) included in (\ref{losshop}) 
weakly influence the power $P$ it can be calculated using equation (\ref{njunja1}) which includes only the lowest order process. 
Therefore the formula for the dissipated power can be rewritten as
\begin{eqnarray}
P=\frac{\hbar\omega_0\rho^2_0}{4\pi}\ \int\ Q^3dQ e^{-2Qa}\hspace{2cm}
\label{njunja2}\\
\nonumber\\
\hspace{2cm}\int^{\omega_0}_{0}\frac{d\omega}{2\pi}\ ImD_1(Q,\omega)ImD_2(Q,\omega_0-\omega).
\nonumber
\end{eqnarray}
This suggests that the dissipated power, within the parameter space used in this 
investigation (for $x<x_{cut}$), behaves as $P\sim\rho^2_0$. Also Eq.\ref{njunja2} 
suggests that the resonant condition (maximum in $P$) will occur when the 
driving frequencies satisfy the condition
\begin{equation}
\omega_0=n_1\omega_i+n_2\omega_j;\ \ n_1,n_2=1,2,3,...     
\end{equation} 
where $\omega_i=\omega_p$, $\omega_{TO1}$ and $\omega_{TO2}$ are the frequencies of hybridized Dirac plasmons and TO phonons, respectively.

\begin{figure*}[tt]
\includegraphics[width=1.0\columnwidth]{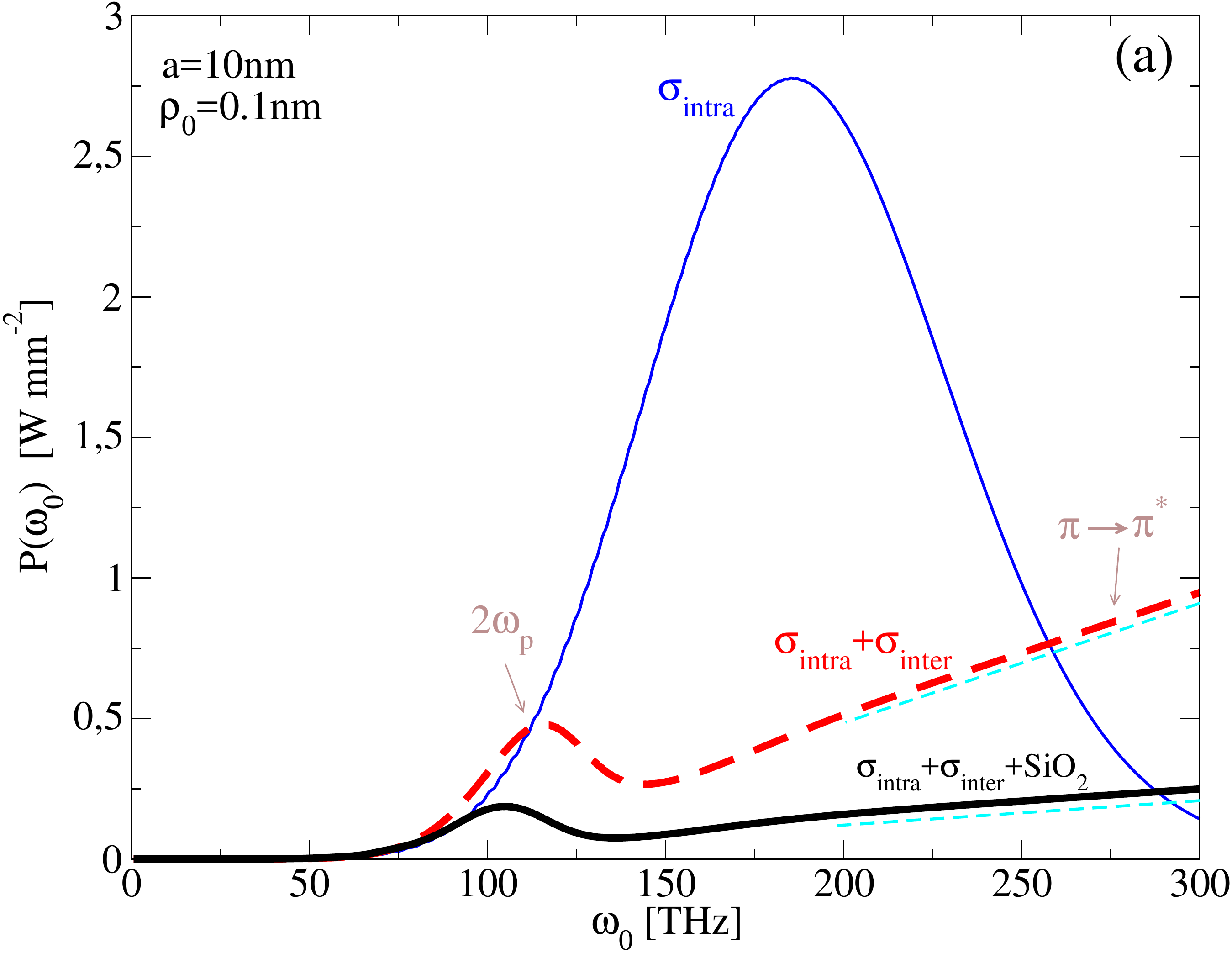}
\includegraphics[width=1.0\columnwidth]{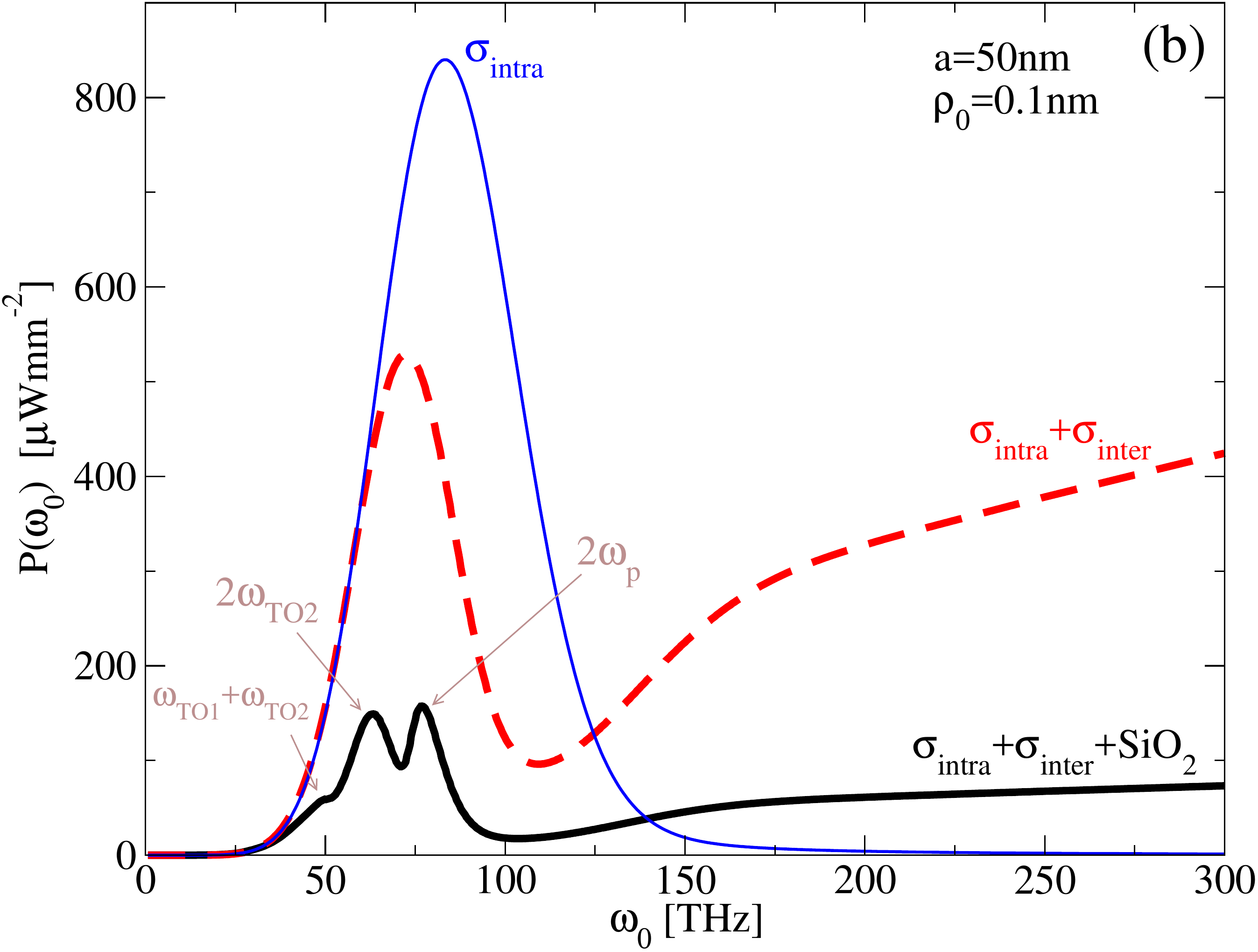}
\includegraphics[width=1.0\columnwidth]{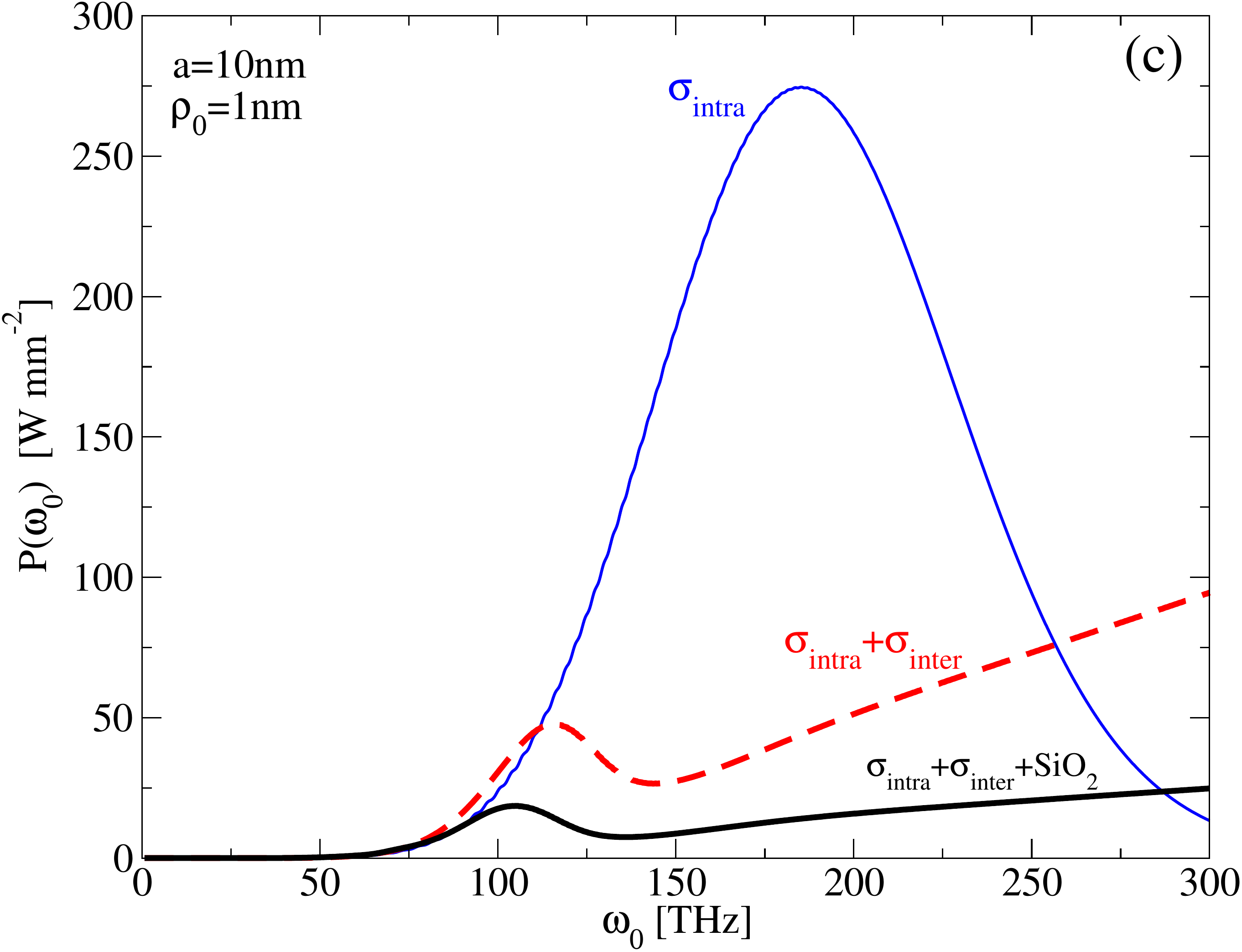}
\includegraphics[width=1.0\columnwidth]{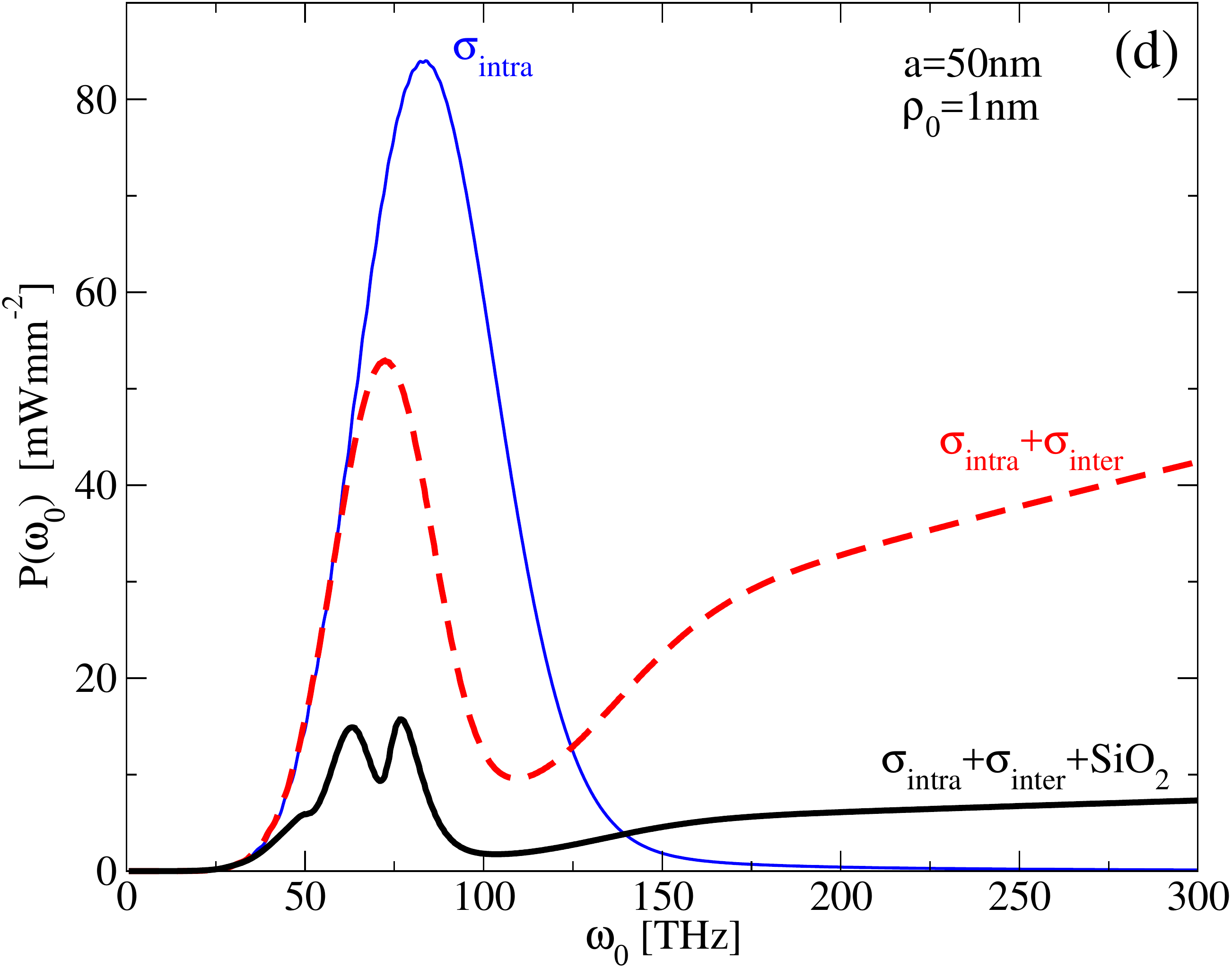}
\caption{(Color online). Dissipated power in two oscillating unsupported Drude 
(\ref{curren1},\ref{curren2}) graphenes (blue thin line), unsupported full conductivity 
(\ref{curren1},\ref{curren2},\ref{curren4}) graphenes (red dashed line) and full conductivity graphenes deposited on semiinfinite ($\Delta\rightarrow\infty$) SiO$_2$ substrates (black solid line). 
The separations between slabs and oscillation amplitudes are (a) $a=10$nm, $\rho_0=0.1$nm, (b) $a=50$nm, $\rho_0=0.1$nm, (c) $a=10$nm, $\rho_0=1$nm and (d) $a=50$nm, $\rho_0=1$nm. Both graphenes are doped such that $E_{F1}=E_{F2}=200$meV.} 
\label{Fig6}
\end{figure*}

Figs \ref{Fig6} show the dissipated power  $P(\omega_0)$ for two oscillating graphene 
monolayers, calculated in several approximations: unsupported graphene (no substrate) 
using Drude expression (\ref{curren1}--\ref{curren2}) for the conductivity (blue thin line), and using full conductivity (\ref{curren1}--\ref{curren2},\ref{curren4}) (red dashed line), as well as for graphenes on semiinfinite ($\Delta\rightarrow\infty$) SiO$_2$ substrates with full expression for 
conductivity (black solid line). Both graphene monolayers are doped so that $E_{F1}=E_{F2}=200$meV. 
In Fig.\ref{Fig6}(a) the separation between slabs and oscillation amplitude are 
$a=10$nm and $\rho_0=0.1$nm, respectively. We see that in the Drude model $P$ shows a strong 
peak which comes from the excitation of undamped Dirac plasmons. 
In the full conductivity model plasmon peak is strongly suppressed and 
interband $\pi\rightarrow\pi^*$ excitations become the dominant dissipation mechanism. The 
fingerprint of $\pi\rightarrow\pi^*$ excitations in Fig.\ref{Fig6}(a) is 
linear $P(\omega_0)$ behaviour starting at about $200$THz, where we also added cyan dashed lines to guide the eye. It can also be noted that the plasmon peak is red shifted which is reasonable considering that $\pi\rightarrow\pi^*$ transitions push Dirac plasmon dispersion toward lower energies.

In the presence of the substrate  dissipation is additionally reduced by almost a 
factor of three. This is because for smaller separations ($a=10$nm) the modes with higher wave 
vectors (e.g. $Q\approx 0.01a.u.$), which is in this case only the Dirac 
plasmon, dominantly contribute to $P$. In this wavevector region the Dirac 
plasmon already has high enough frequency ($\omega\approx 60$THz) that the dynamical part 
of the substrate screening in not active and the substrate dielectric function can be 
approximated by $\epsilon_S(\omega)\approx\epsilon_{\infty}$. This causes the reduction of substrate screened Coulomb interaction $\tilde{v}_Q(\omega)=\frac{2}{1+\epsilon_S(\omega)}v_Q$ 
(see Eq.\ref{tildeV11}) and then (considering Eq.\ref{newD}) reduction of the plasmon intensity, which finally causes the reduction of $P$. Reduction of the screened Coulomb interaction (\ref{tildeV11}) also causes the reduction of the plasmon frequency which can also be noted.

Fig.\ref{Fig6}(b) shows the dissipated power $P(\omega_0)$ for the same set of parameters as 
in Fig.\ref{Fig6}(a) except that the separation between slabs is increased to $a=50nm$.
As expected, from the discussion in Sec.\ref{weakvdW}, $P$ is reduced by about four 
orders of magnitude and plasmon peaks are shifted toward lower frequencies. The latter is also expected considering that for larger separations the modes with smaller $Q$ contribute, and here 
the Dirac plasmon has lower energy. We can notice qualitative difference between $P$ in Figs.\ref{Fig6}(a) and (b) for the case when substrate is present (black lines). 
In Fig.\ref{Fig6}(b) $P$ possesses additional structures (two additional peaks at 
$\omega_{TO1}+\omega_{TO2}$ and $2\omega_{TO2}$) which are not present in Fig.\ref{Fig6}(a). This is because for larger $a$ the modes with smaller wave vectors (e.g. $Q\approx 0.002$a.u.) 
start contributing to $P$, and this is exactly the region where plasmon/phonon hybridization 
occurs (as ilustrated in Fig.\ref{Fig4}(a)), so the additional peaks at $\omega_{TO1}+\omega_{TO2}$ 
and $2\omega_{TO2}$ represent the resonant dissipation to two phonon modes.

Figs.\ref{Fig6}(c) and (d) show the dissipated power $P$ for the same parameters 
as in Figs.\ref{Fig6}(a) and (b), respectively, except that the oscillation amplitude is 
increased to $\rho_0=1$nm. $P$ in Figs.\ref{Fig6}(c) 
and (d) are qualitatively the same and exactly hundred times 
larger than $P$ in Figs.\ref{Fig6}(a) and (b). This confirms $P\sim\rho^2_0$ 
behaviour of the dissipated power with amplitude as predicted by Eq.\ref{njunja2}. 

\subsection{Dissipated power - graphene doping and distance dependence}
\label{DISdop}

In this section we shall explore the dissipated power for two 
oscillationg slabs for different graphene dopings.   

\begin{figure*}[tt]
\includegraphics[width=1.0\columnwidth]{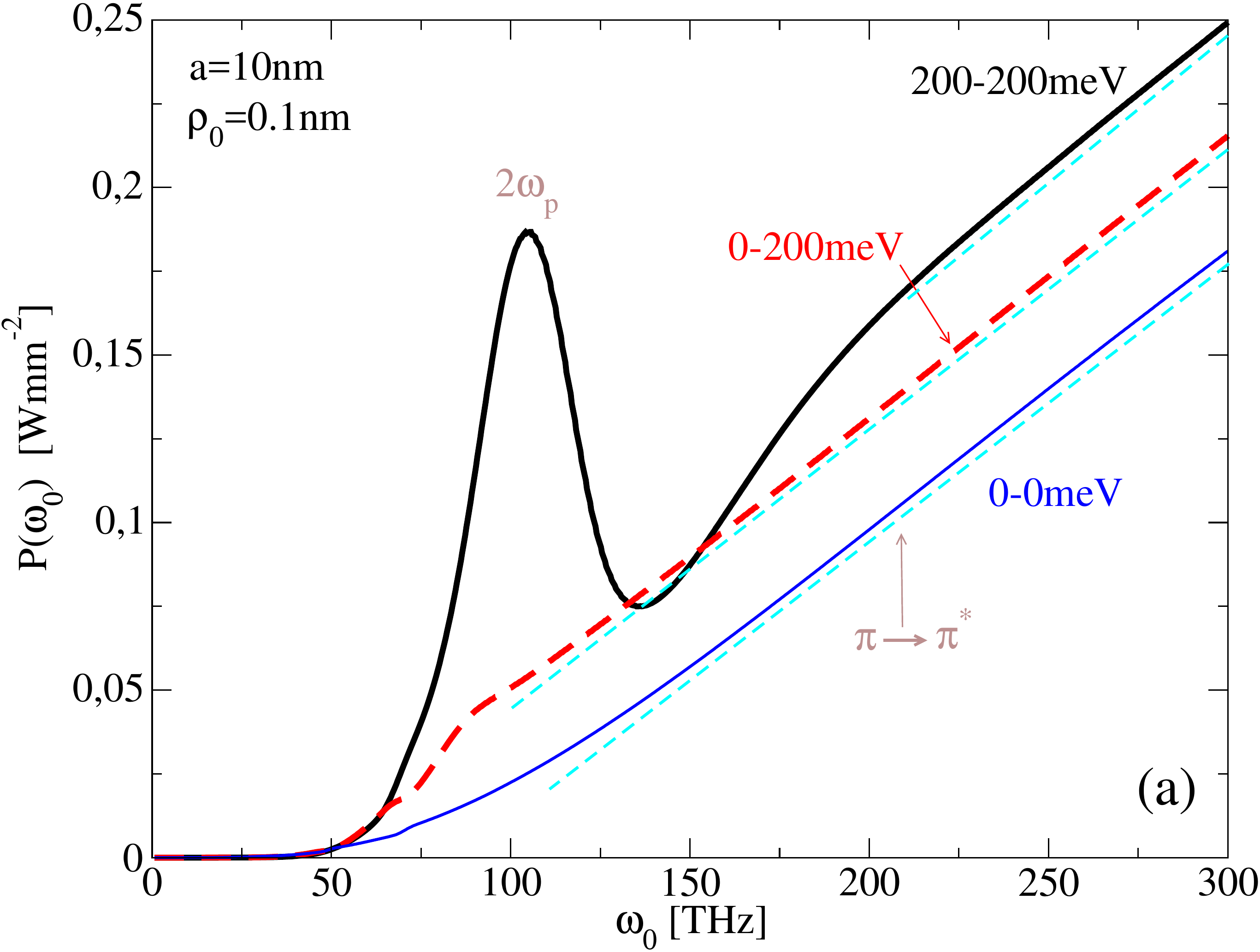}
\includegraphics[width=1.0\columnwidth]{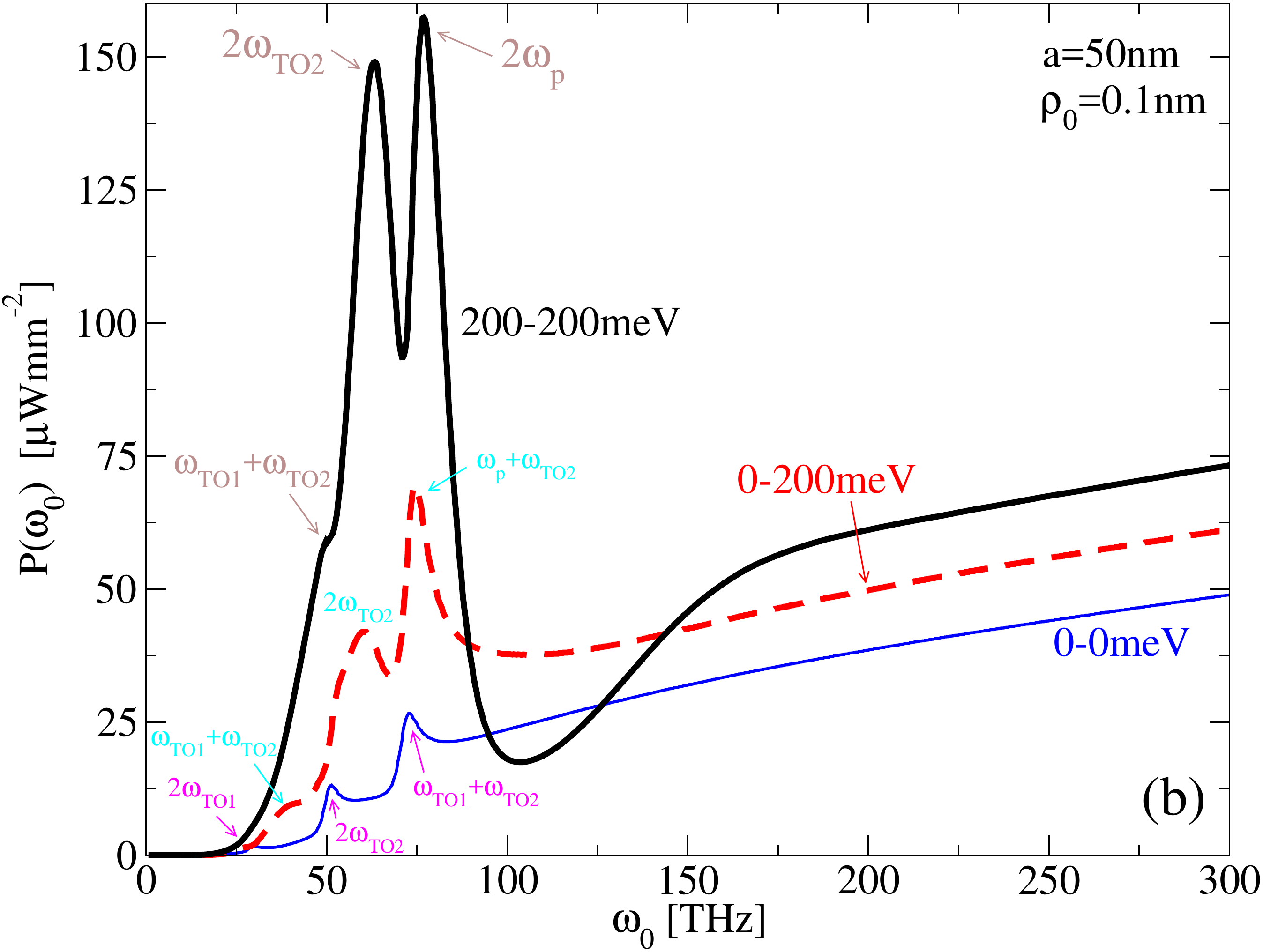}
\includegraphics[width=1.0\columnwidth]{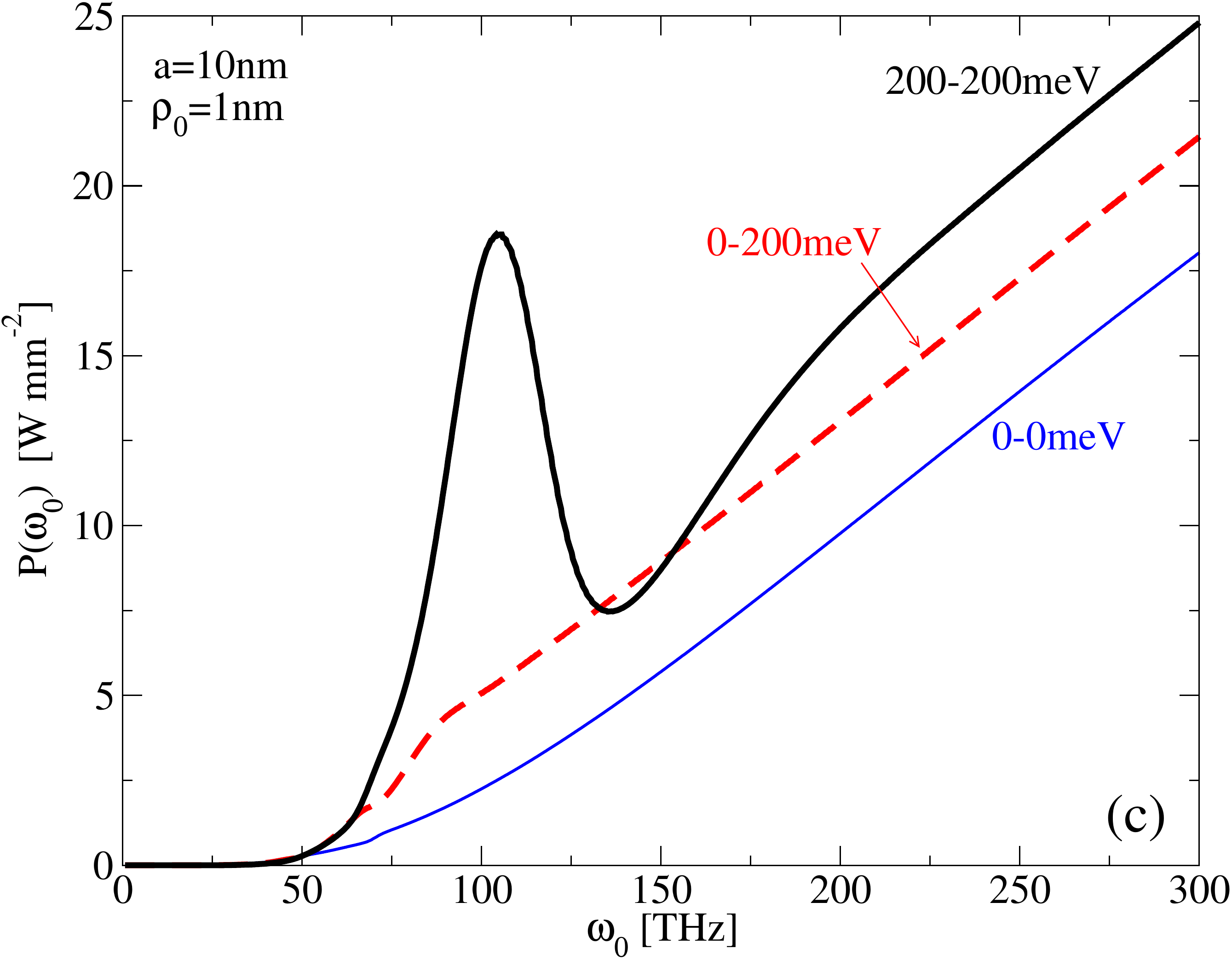}
\includegraphics[width=1.0\columnwidth]{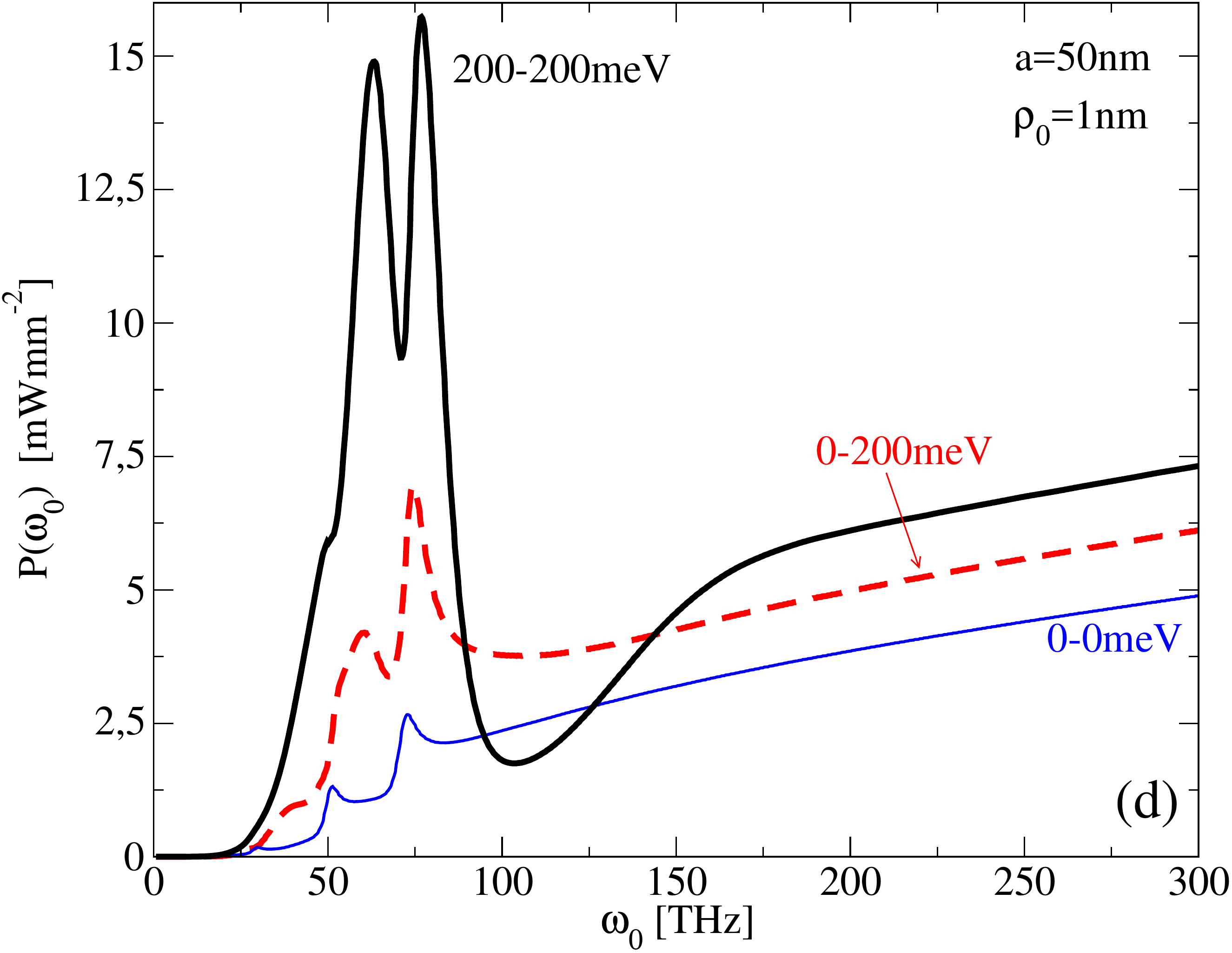}
\caption{(Color online). Dissipated power for two oscillating graphenes deposited on semiinfinite ($\Delta\rightarrow\infty$) SiO$_2$ substrates where the graphene dopings $E_{F1}-E_{F2}$ are $0-0$meV (blue thin line) $0-200$meV (red dashed line) and  $200-200$meV (black solid line). The separations between slabs and oscillation amplitudes are (a) $a=10$nm, $\rho_0=0.1$nm, (b) $a=50$nm, $\rho_0=0.1$nm, (c) $a=10$nm, $\rho_0=1$nm and (d) $a=50$nm, $\rho_0=1$nm. The graphene response is 
calculated using full conductivity expression (\ref{curren1}--\ref{chi-sigma}).} 
\label{Fig7}
\end{figure*}

Fig.\ref{Fig7}(a) shows the dissipated power in two oscillating graphenes deposited 
on semiinfinite ($\Delta\rightarrow\infty$) SiO$_2$ substrates where the graphene 
dopings $E_{F1}-E_{F2}$ are $0-0$meV (blue thin line) $0-200$meV (red dashed line) and 
$200-200$meV (black solid line). The separations between slabs and oscillation amplitude 
are $a=10$nm and $\rho_0=0.1$nm, respectively. 

If both graphenes are doped $P$ shows the plasmon peak at 
about $2\omega_p=100$THz, and starting at about $200$THz it increases 
linearly, which is the consequence of interband $\pi\rightarrow\pi^*$ excitations, as 
already observed in Fig.\ref{Fig6}. 
However, if one doped graphene is replaced by pristine graphene ($E_F=0$), which does 
not support the Dirac plasmon (as shown in Fig.\ref{Fig4}(c)), the Dirac plasmon in doped 
graphene can no longer resonantly transfer energy to the Dirac plasmon in another graphene 
and $P$ loses the plasmon peak at $2\omega_p$. 
However, the visible step remains (at about $75$THz) which is the consequence of energy 
transfer between Dirac plasmon in the doped graphene and $\pi\rightarrow\pi^*$ excitations 
in the undoped one. In this case (small $a$ and larger $Q$) phonons are still very weak 
and do not represent important dissipation channel. When both graphenes are pristine the 
only dissipation comes from the resonant energy transfer between $\pi\rightarrow\pi^*$ 
excitations in different graphenes, resulting in the strictly linear behaviour of 
$P$.  

Fig.\ref{Fig7}(b) shows the dissipated power $P$ for the same parameters as in 
Fig.\ref{Fig7}(a) except that the separation between slabs is increased 
to $a=50nm$. As we have already discussed in Fig.\ref{Fig6}(a), in this case the modes with 
smaller wave vectors $Q$ contribute and the dissipated power $P$ gets additional 
structures coming from resonant phonon excitations. For the case $E_{F1}-E_{F2}=200-200$meV (black solid line) (coupling between modes in Fig.\ref{Fig4}a) the dissipated power shows three 
peaks at $\omega_{TO1}+\omega_{TO2}\approx 40$THz, $2\omega_{TO2}\approx 60$THz and 
$2\omega_{p}\approx 75$THz. For the case $E_{F1}-E_{F2}=0-200$meV (red dashed line) there 
is a possibility for resonant coupling between two phonons in the slab with pristine graphene and 
three hybridized plasmon/phonon modes in the slab with doped graphene (coupling beteen modes in Fig.\ref{Fig4}a and modes in Fig.\ref{Fig4}c). The three peaks correspond to resonant 
couplings at $\omega_{TO1}+\omega_{TO2}$, $2\omega_{TO2}$ and $\omega_{TO2}+\omega_p$, 
as denoted in Fig.\ref{Fig7}(b). When both graphenes are pristine, i.e. 
$E_{F1}-E_{F2}=0-0$meV (thin solid blue line) the dominant dissipation channels 
become the resonant coupling between phonons in both slabs (coupling between 
modes in Fig.\ref{Fig4}(c)). The three peaks correspond to resonant couplings at $2\omega_{TO1}$, $\omega_{TO1}+\omega_{TO2}$ and $2\omega_{TO2}$, as denoted in Fig.\ref{Fig7}(b). 
Of course, in all three cases $P$ shows linear behaviour for larger 
$\omega_0$ coming from the resonant $\pi\rightarrow\pi^*$ excitations in both slabs. 
Figs.\ref{Fig7}(c) and (d) show the same as Figs.\ref{Fig7} (a) and (b), except 
that the oscillation amplitude is increased to $\rho_0=1$nm.
As in Figs.\ref{Fig6}, $P$ is qualitatively similar and exactly 
hundred times larger than $P$ in Figs.\ref{Fig7}(a) and (b). 
This again confirms the $P\sim\rho^2_0$ behaviour. 

This strong dependence of dissipated power on graphene 
doping suggests many opportunities for applications.

\section{Conclusions}
\label{Sec5}

In this paper we have provided a complete theoretical description of van der Waals and friction 
forces for two slabs in relative oscillatory motion which 
includes variable temperatures in two slabs, their dynamical properties, and dependence on characteristic oscillation amplitude 
and frequency. In Appendix we also provide, for comparison, analogous expressions for the slabs in parallel uniform motion. 

We applied this formulation to explore van der Waals and friction forces between two oscillating slabs, each consisting of atomically thick 
crystal (e.g. graphene) adsorbed on a dielectric substrate (SiO$_2$). 
We explore dependence of these forces on  osillator characteristics such  as driving amplitude 
$\rho_0$ and frequeny $\omega_0$, but also on slab separation $a$, on graphene doping $E_F$ and on substrate properties.   
We show how the spectra of coupled electronic/phononic excitations in slabs determine the energy transfer 
processes in this system. 

We show that, in general, as the driving frequency $\omega_0$ increases the vdW energy first shows an unusual plateau, and then decreases. 
We propose the idea of controlling  the 'sticking' and 'un-sticking' of slabs by tuning the 
graphene dopings $E_{Fi}$ and driving frequency $\omega_0$.

We also found a simple $\rho^2_0$ dependence of both the vdW force  and dissipated power. 
The dissipated power between Drude model graphenes, as function of $\omega_0$, shows  unrealistically strong 
$2\omega_p$ peak. However, in a realistic graphene (whose dielectric properties are calculated 
from first principles) this peak is strongly reduced and red shifted. We also explain why the substrate substantially reduces dissipated power peak $2\omega_p$.  
For larger separations $a$ additional peaks appears in dissipation power originating from the excitations of hybridized substrate phonons.     

We showed that if one graphene is pristine ($E_F=0$) it causes the disappearance of the strong $2\omega_p$ 
peak in the dissipated power. Moreover, for larger separations $a$ the doping causes shifts, appearance  and disappearance  of many peaks 
originating from resonant coupling between hybridised electronic/phononic excitations in graphene/substrate slabs.

Obviously, when present, the Dirac plasmons  are the dominant channels through which the energy between slabs can be 
transferred, so  the studied model system strongly supports the possibility to control the energy or heat transfer between 
the slabs by tuning the graphene doping, e.g. by  electrostatic gating.
More radically, for zero doping $E_F=0$ the energy transfer can be locked, and vice 
versa.    

In conclusion, it is expected that studies of energy transfer processes in the case of 
osillating slabs, based on our complete theoretical description, will provide supplementary and more 
practical approach as compared to those in parallel uniform motion.

\section*{Acknowledgments}
Two of the authors (V. D. and M. \v S.) are grateful for the hospitality at the Donostia International Physics 
Center where this work was finalized, and for useful discussions to J. Pendry, A. A. Lucas, S. Silkin  and I. Kup\v ci\' c. V. D. acknowledges the support of the University of the Basque Country and the Spanish Ministerio de Ciencia y Tehnologia.
V. D. also acknowledges the support of QuantiXLie Centre of Excellence, a project
cofinanced by the Croatian Government and European Union through the
European Regional Development Fund - the Competitiveness and Cohesion
Operational Programme (Grant KK.01.1.1.01.0004).

\appendix
\numberwithin{equation}{section}
\section{General theory-Uniformly moving slabs}
\label{AppA}
\subsection{Van der Waals energy and force}
\label{vdWenergyforce}
We shall first derive the van der Waals potential and force between two inequivalent slabs, described 
by their response functions $R_1$ and $R_2$, moving with relative parallel velocity  ${\bf v}$ and separated 
by $a$, as can be seen in Fig.\ref{FigA1}. In the following we shall briefly summarize the derivation presented in 
Ref.\cite{trenjePRB}, modified to describe the most general case, i.e. for the slabs with different response functions $R_1\ne R_2$  
and different temperatures $T_1\ne T_2$, including the case of graphene monolayers deposited on dielectric substrates.  
In the diagram in Fig.\ref{Fig1}b the density fluctuation $S_1$ in the slab 1 
couples to the induced charge described by $D_2$ inside the slab 2, which results in the ground state energy shift:
\begin{figure}[h]
\centering
\includegraphics[width=7.0cm,height=4cm]{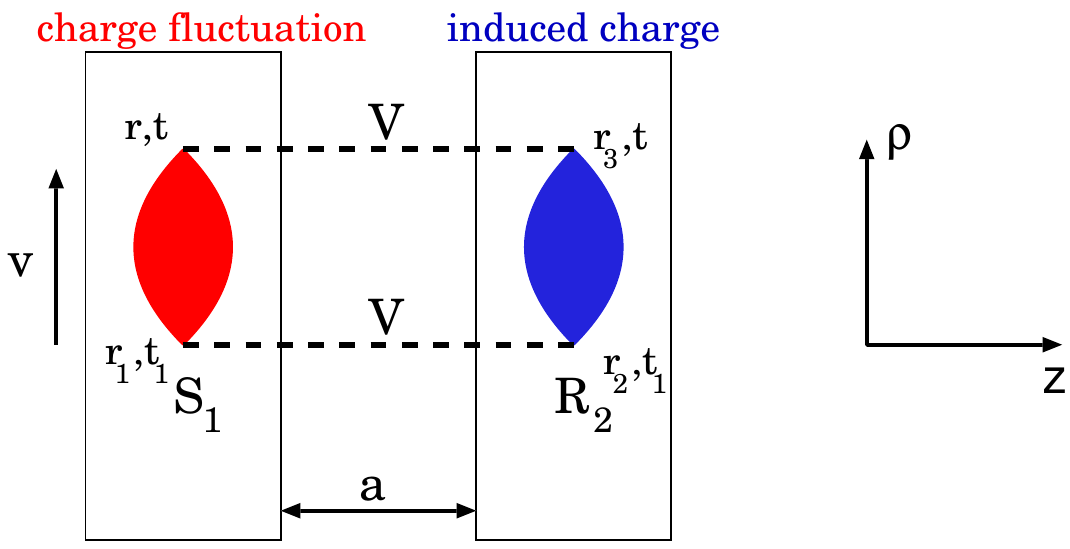}
\caption{Process in which charge density fluctuation is created in the left slab and induces a potential 
in the right slab.} 
\label{FigA1}
\end{figure}

\begin{eqnarray}
E_{c}=\int^{\infty}_{-\infty}dt_1\int d{\bf r}\int d{\bf r}_1\int d{\bf r}_2\int d{\bf r}_3 
\nonumber\\
\nonumber\\
S_1({\bf r},{\bf r}_1,t,t_1)V({\bf r},{\bf r}_3)R_2({\bf r}_3,{\bf r}_2,t,t_1)V({\bf r}_2,{\bf r}_1)  
\label{jujujuju}
\end{eqnarray}
Here $V$ is the Coulomb potential, $S_1$ is the correlation function of the left 
slab and $R_2$ is the response function of the right slab. We assume that the slab 1 is moving with velocity ${\bf v}$ so that 
the parallel coordinates in $S_1$ are transformed 
as 
\begin{equation}
\brho-\brho_1\rightarrow\brho-\brho_1-{\bf v}(t-t_1).
\label{njuski}
\end{equation}
If we use translational invariance in time and in the parallel direction and perform the Fourier transform in parallel 
coordinates we find    
\begin{eqnarray}
E_{c}=\int^{\infty}_{-\infty}dt_1\int\frac{d{\bf Q}}{(2\pi)^2}
e^{-i{\bf Q}{\bf v}(t-t_1)}
\nonumber\\
\int^{\infty}_{-\infty} dzdz_1dz_2dz_3 
S_1({\bf Q},z,z_1,t-t_1)V({\bf Q},z,z_3)\times
\nonumber\\
R_2({\bf Q},z_3,z_2,t-t_1)V({\bf Q},z_2,z_1).\hspace{2cm}
\label{nabij3}
\end{eqnarray}
The Fourier transform in time gives: 
\begin{eqnarray}
E_{c}=\int^{\infty}_{-\infty}\frac{d\omega}{2\pi}\int\frac{d{\bf Q}}{(2\pi)^2}
\int^{\infty}_{-\infty} dzdz_1dz_2dz_3\hspace{2cm}
\label{jujujuju1}\\
S_1({\bf Q},z,z_1,\left|\omega\right|)V({\bf Q},z,z_3)R_2({\bf Q},z_3,z_2,\omega')V({\bf Q},z_2,z_1),
\nonumber
\end{eqnarray} 
where we have introduced $\omega'=\omega+{\bf Q}{\bf v}$.
Because the charge densities in slabs 1 and 2 do not overlap, $z$ integrals in (\ref{jujujuju1}) contribute only for 
$z_3>z$ and $z_2>z_1$, so that we can write 
\begin{eqnarray}
V({\bf Q},z,z_3)=v_Qe^{-Q(z_3-z)},
\nonumber\\
\nonumber\\
V({\bf Q},z_2,z_1)=v_Qe^{-Q(z_2-z_1)}.
\label{Def1}
\end{eqnarray}
where $v_Q=\frac{2\pi e^2}{Q}$.
Also, if we use the definition of the surface correlation function
\begin{equation}
S_1({\bf Q},|\omega|)=v_Q\int dzdz_1e^{Qz}S_1({\bf Q},|\omega|,z,z_1)e^{Qz_1},
\label{Def2}
\end{equation}
and the definition of the surface excitation propagator \cite{PhysScr,JPCM} 
\begin{equation}
D_2({\bf Q},\omega)=v_Q\int dz_2dz_3e^{Qz_2}R_2({\bf Q},\omega,z_2,z_3)e^{Qz_3},
\label{Def3}
\end{equation}
expression (\ref{jujujuju1}) can be written as  
\begin{eqnarray}
E_{c}=\int^{\infty}_{-\infty}\frac{d\omega}{2\pi}\int\frac{d{\bf Q}}{(2\pi)^2}\ e^{-2Qa}\hspace{2cm}
\nonumber\\
\nonumber\\
S_1({\bf Q},|\omega|)ReD_2({\bf Q},\omega')  
\label{jujujuju2}
\end{eqnarray} 
Moreover, after we use the connection between the thermal/quantum mechanical charge density fluctuations and the 
dissipation in the left slab:  
\begin{equation}
S_1({\bf Q},\left|\omega\right|)=-\frac{\hbar}{\pi}\ [2n_1(\omega)+1]ImD_1({\bf Q},\omega),  
\label{impsv}
\end{equation}
where $n_1(\omega)=1/(e^{\beta_1\hbar\omega}-1)$ represents the Bose-Einstein distribution, $\beta=k_BT_1$ and $T_1$ is the temperture of slab 1, 
the expression (\ref{jujujuju2}) becomes
\begin{eqnarray}
E_{c}=-\frac{\hbar}{\pi}\int^{\infty}_{-\infty}\frac{d\omega}{2\pi}[2n_1(\omega)+1]\int\frac{d{\bf Q}}{(2\pi)^2}\ e^{-2Qa}
\nonumber\\
\nonumber\\
ImD_1({\bf Q},\omega)ReD_2({\bf Q},\omega').\hspace{2cm}   
\label{jujujuju3}
\end{eqnarray} 
Here we have used the fact that $ImD_2({\bf Q},\omega)$ is an odd function of $\omega$ and 
does not contribute in (\ref{jujujuju2}).     
To this we have to add the contribution from the process in which the charge density 
fluctuation is created  in the slab 2. Because then slab $2$ moves with parallel velocity {\bf v} 
relative to slab 1 this contribution can be obtained from (\ref{jujujuju3}) by 
exchanging ${\bf v}\rightarrow-{\bf v}$ and $1\leftrightarrow 2$, and the result for the 
van der Waals energy is:
\begin{eqnarray}
E_{c}=-\frac{\hbar}{\pi}\int\frac{d{\bf Q}}{(2\pi)^2}e^{-2Qa}\int^{\infty}_{-\infty}\frac{d\omega}{2\pi}\times\hspace{2cm}
\label{jura66}\\
\nonumber\\
\left\{[2n_1(\omega)+1]ImD_1({\bf Q},\omega)ReD_2({\bf Q},\omega')+\right.
\nonumber\\
\nonumber\\
\left.[2n_2(\omega)+1]
ImD_2({\bf Q},\omega)ReD_1({\bf Q},\omega')
\right\}.
\nonumber
\end{eqnarray} 
$E_c$ given by (\ref{jura66}) includes only the lowest order processes shown in Fig.\ref{Fig1}a. 
If we want to include higher order processes shown in Fig.\ref{FigA2}, we have to replace 
the interaction $v_Q$ which appears in $D_1$:  
\newline
\begin{equation}
v_Q\ \rightarrow\  v_Q(1+D_2D_1e^{-2Qa}+...)=\frac{v_Q}{1-D_2D_1e^{-2Qa}}
\label{Eka1}
\end{equation}
and the one which appears in $D_2$:
\begin{equation}
v_Q\ \rightarrow\  v_Q(1+D_1^*D^*_2e^{-2Qa}+...)=\frac{v_Q}{1-D_1^*D^*_2e^{-2Qa}}
\label{Eka2}
\end{equation}
and integrate over the coupling constant $\lambda$ to find 
\begin{eqnarray}
E_{c}=\hbar\int^{1}_0\frac{d\lambda}{\lambda}\int\frac{d{\bf Q}}{(2\pi)^2}e^{-2Qa}
\int^{\infty}_{-\infty}\frac{d\omega}{2\pi}[2n_1(\omega)+1]
\nonumber\\
\left[\frac{\lambda^2ImD_1({\bf Q},\omega)ReD_2({\bf Q},\omega')}{\left|1-\lambda^2e^{-2Qa}D_1({\bf Q},\omega)D_2({\bf Q},\omega')\right|^2}+
\left(
1\leftrightarrow 2
\right)
\right].\hspace{1cm}
\label{jujujuju5}
\end{eqnarray}
\begin{figure}[h]
\centering
\includegraphics[width=1.0\columnwidth]{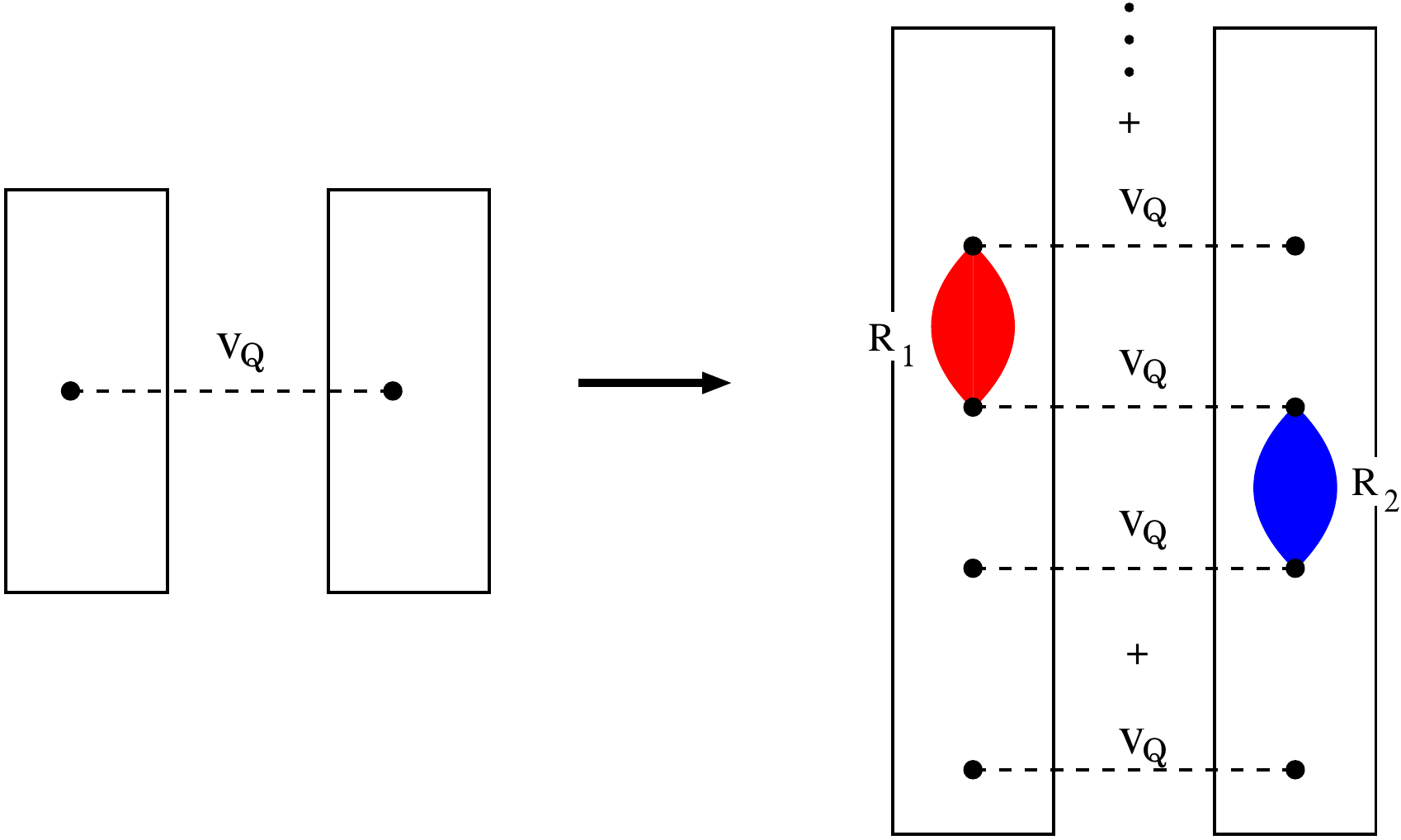}
\caption{Higher order processes.} 
\label{FigA2}
\end{figure}
Notice that (\ref{jujujuju5}) does not change for ${\bf v}\rightarrow-{\bf v}$.
In order to do the $\lambda$ integration we transform this expression into:   
\begin{widetext}
\begin{equation}
E_{c}=\hbar\int^{1}_0\frac{d\lambda}{\lambda}\int\frac{d{\bf Q}}{(2\pi)^2}
\int^{\infty}_{-\infty}\frac{d\omega}{2\pi}[2n_1(\omega)+1]
\left\{
\frac{ImD_1({\bf Q},\omega)ReD_2({\bf Q},\omega')}{Im\left[D_1({\bf Q},\omega)D_2({\bf Q},\omega')\right]}
Im\frac{\lambda^2e^{-2Qa}D_1({\bf Q},\omega)D_2({\bf Q},\omega')}{1-\lambda^2e^{-2Qa}D_1({\bf Q},\omega)D_2({\bf Q},\omega')}+
\left(
1\leftrightarrow 2
\right)
\right\}
\label{jujujuju6}
\end{equation}
\end{widetext}
which finally gives the van der Waals energy in the case of unequal slabs and finite velocity:
\begin{equation}
E_{c}(a)=\frac{\hbar}{2}\int\frac{d{\bf Q}}{(2\pi)^2}
\int^{\infty}_{-\infty}\frac{d\omega}{2\pi}\ A({\bf Q},\omega,\omega')
\label{jujujuju7}
\end{equation}
where 
\begin{equation}
A({\bf Q},\omega,\omega')=[2n_1(\omega)+1]A_{12}({\bf Q},\omega,\omega')+(1\leftrightarrow 2)
\label{jalko}
\end{equation}
and 
\begin{eqnarray}
A_{ij}({\bf Q},\omega,\omega')=
\frac{ImD_i({\bf Q},\omega)ReD_j({\bf Q},\omega')}{Im\left[D_i({\bf Q},\omega)D_j({\bf Q},\omega')\right]}\times\hspace{1cm}
\nonumber\\
\nonumber\\
Im\ln\left[1-e^{-2Qa}D_i({\bf Q},\omega)D_j({\bf Q},\omega')\right].\hspace{1cm}
\label{jujucka}
\end{eqnarray}
One can easily derive the limiting cases. For $T_1=T_2=0$ $2n_1(\omega)+1\rightarrow sgn\omega$. For ${\bf v}=0$ and $\omega'=\omega$ the 
expression (\ref{jalko}) becomes (for $T=0$):   
\begin{equation}
A({\bf Q},\omega=\omega')=Im\ln\left[1-e^{-2Qa}D_1({\bf Q},\omega)D_2({\bf Q},\omega)\right].
\end{equation}
For ${\bf v}\ne0$ but $D_1=D_2=D$ it becomes: 
\begin{eqnarray}
A({\bf Q},\omega,\omega')=2\ \frac{ImD({\bf Q},\omega)ReD({\bf Q},\omega')}
{Im\left[D({\bf Q},\omega)D({\bf Q},\omega')\right]}\times
\nonumber\\
Im\ln\left[1-e^{-2Qa}D({\bf Q},\omega)D({\bf Q},\omega')\right].\hspace{2cm}
\label{sinko}
\end{eqnarray} 
From the van der Waals potential $E_c(a)$ we can derive the perpendicular attractive force $F_{\perp}(a)$ between 
two moving slabs: 
\begin{eqnarray}
F_{\perp}(a)=-\frac{dE_{c}(a)}{da}=\hspace{3cm}
\nonumber\\
\nonumber\\
\hbar\int\frac{d{\bf Q}}{(2\pi)^2}Qe^{-2Qa}\int^{\infty}_{-\infty}\frac{d\omega}
{2\pi}\ B({\bf Q},\omega,\omega')
\label{jujujuju99}
\end{eqnarray}
where 
\begin{equation}
B({\bf Q},\omega,\omega')=[2n_1(\omega)+1]B_{12}({\bf Q},\omega,\omega')+(1\rightarrow 2)
\label{prcko}
\end{equation}
and
\begin{equation}
B_{ij}({\bf Q},\omega,\omega')=\frac{ImD_i({\bf Q},\omega)ReD_j({\bf Q},\omega')}{\left|1-e^{-2Qa}D_i({\bf Q},\omega)D_j({\bf Q},\omega')\right|^2}.
\label{prdf99}
\end{equation}
Again for ${\bf v}=0$ and $T_1=T_2=0$ the 
expression (\ref{prcko}) becomes 
\begin{equation}
B({\bf Q},\omega=\omega')=\frac{Im[D_1({\bf Q},\omega)D_2({\bf Q},\omega)]}{\left|1-e^{-2Qa}D_1({\bf Q},\omega)D_2({\bf Q},\omega)\right|^2},
\label{jurec}
\end{equation}
and for ${\bf v}\ne0$ but $D_1=D_2=D$ it becomes 
\begin{equation}
B({\bf Q},\omega,\omega')=2\ \frac{ImD({\bf Q},\omega)ReD({\bf Q},\omega')}{\left|1-e^{-2Qa}D({\bf Q},\omega)D({\bf Q},\omega')\right|^2}.
\label{kurec}
\end{equation}
We note that for ${\bf v}=0$ our results agree with the previous ones, but for ${\bf v}\ne0$ they differ from those in Ref.\cite{Pendry1,Persson1}. 
The functions $A$ and $B$ will also appear in the same form in the expressions for van der Waals potential and 
force between the oscillating slabs, but with the different choice for $\omega'$.

We can verify, using spectral representations for $ReD$'s, that our results correspond exactly to the well known 
result for the van der Waals attraction between two moving or oscillating objects in the lowest order \cite{Pedro}, e.g. for $T=0$:    
\begin{eqnarray}
E^{(2)}_{c}=
\frac{\hbar}{2}\int\frac{d{\bf Q}}{(2\pi)^2}e^{-2Qa}
\int^{\infty}_{-\infty}\frac{d\omega}{2\pi}sgn\omega\hspace{1cm}
\\
\nonumber\\
\left[ImD_1({\bf Q},\omega)ReD_2({\bf Q},\omega')+
ImD_2({\bf Q},\omega)ReD_1({\bf Q},\omega')
\right]
\nonumber
\end{eqnarray}
where $\omega'=\omega+\Delta\omega$ and $\Delta\omega={\bf Q}{\bf v}$  for uniform motion or $\Delta\omega=n\omega_0$ for an oscillator. The frequency 
integral can be rewritten as: 
\begin{eqnarray}
\int^{\infty}_{-\infty}\frac{d\omega}{2\pi}sgn\omega
\left[ImD_1({\bf Q},\omega)ReD_2({\bf Q},\omega')\right.+\hspace{2cm}
\\
\nonumber\\
\left.ImD_2({\bf Q},\omega)ReD_1({\bf Q},\omega')
\right]=
\nonumber\\
\nonumber\\
4\int^{\infty}_{0}d\omega d\nu\frac{ImD_1({\bf Q},\omega)ImD_2({\bf Q},\nu)}{\omega+\nu+\Delta\omega}.
\nonumber
\end{eqnarray}
which is exactly the lowest order term in (\ref{jujujuju7}). 
\subsection{Dissipated power and friction force}
\label{jurniga}

Now we shall calculate the energy dissipated by the two slabs in parallel uniform 
motion following the derivation in Ref.\cite{trenjePRB}.   

\begin{figure}[h]
\centering
\includegraphics[width=9cm,height=3cm]{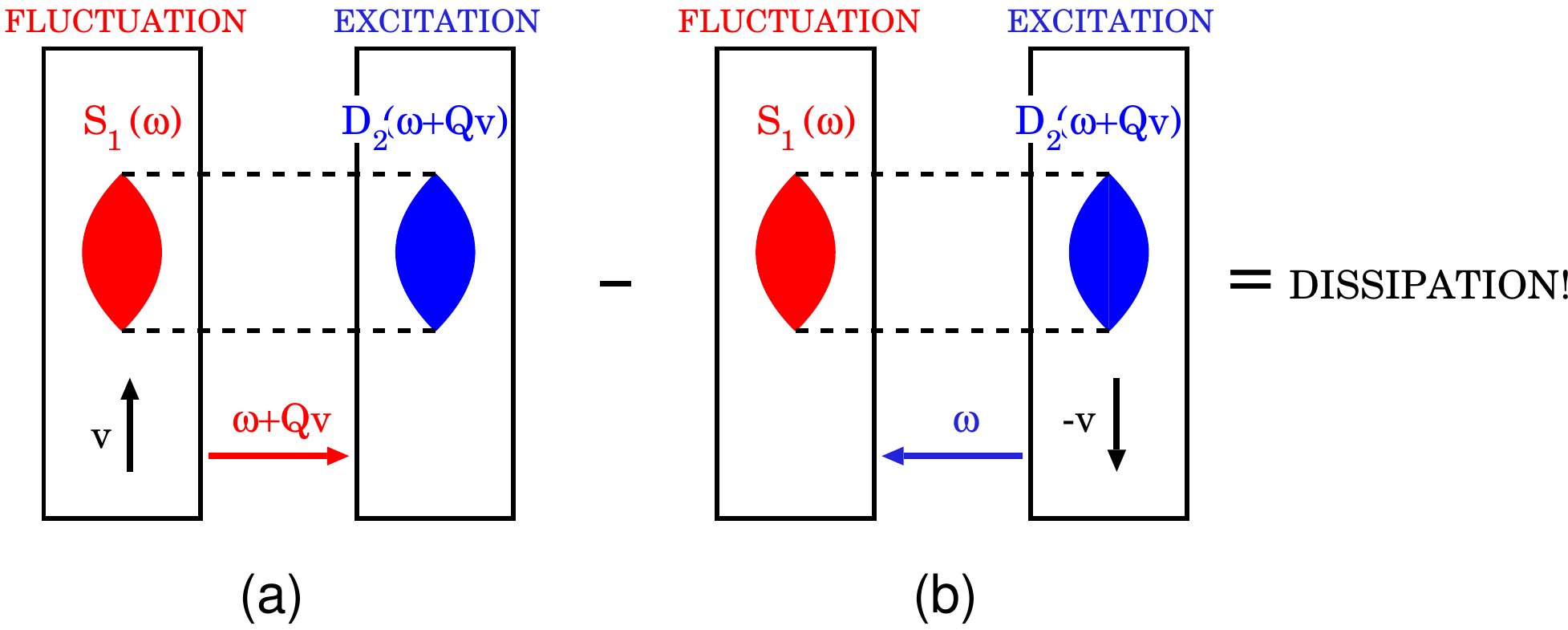}
\caption{Process in which the energy $\omega+{\bf v}{\bf Q}$ is transferred from the left to 
the right slab (a), and the reverse process in which energy $\omega$ is returned 
back to the left slab (b).}
\label{FigA3}
\end{figure}

Suppose that the left slab is moving parallel to the right one with relative velocity ${\bf v}$ and that 
a charge density fluctuation is spontaneously created in the left slab at the moment $t_1$ (Fig.\ref{Fig1}). 
Propagating in time between $t_1$ and $t$ it induces charge density fluctuations in the right slab with which it can subsequently interact.      
In such a process the left slab can be considered as a source which is transferring energy to the right slab, and in analogy with Eqs.3 and 4 
of Ref.\cite{trenjePRB}, the energy loss rate operator in this process can be written 
as 
\begin{equation}
\begin{array}{c}
\hat{P}_{12}=\int d{\bf r}\int d{\bf r}_1\int^{\infty}_{-\infty} dt_1\ \hat{\rho}({\bf r},t)V({\bf r},{\bf r}_3)\otimes 
\\
\frac{d}{dt}D_2({\bf r}_3,{\bf r}_2,t,t_1)\otimes
V({\bf r}_2,{\bf r}_1)\hat{\rho}({\bf r}_1,t_1) 
\end{array}
\label{losse4}
\end{equation}
where $D_2$ is the retarded response function of the right slab and $\hat{\rho}({\bf r},t)$ and $\hat{\rho}({\bf r},t)$ are density operators 
which represent quantum mechanical charge density fluctuations created and annihilated at points $({\bf r}_1,t_1)$  and $({\bf r},t)$, respectively.
Energy transfer rate from the left to the right slab can be obtained by taking the ground state matrix element of Eq.(\ref{losse4})
\begin{eqnarray}
P_{12}=\left\langle\hat{P}_{12}(t)\right\rangle=\int d{\bf r}\int d{\bf r}_1 \int^{\infty}_{-\infty} dt_1\hspace{2cm}
\nonumber\\
\label{losse5}\\
S_1({\bf r},{\bf r}_1,t,t_1)V({\bf r},{\bf r}_3)\otimes
\frac{d}{dt}D_2({\bf r}_3,{\bf r}_2,t,t_1)\otimes V({\bf r}_2,{\bf r}_1)  
\nonumber
\end{eqnarray}
where 
\begin{equation}
S_1({\bf r},{\bf r}_1,t,t_1)=\left\langle\hat{\rho}({\bf r},t)\hat{\rho}({\bf r}_1,t_1) \right\rangle+\left\langle\hat{\rho}({\bf r}_1,t_1)\hat{\rho}({\bf r},t) \right\rangle
\label{correlajme}
\end{equation}
is the correlation function of the left slab which represents real charge density fluctuation.
Eq. (\ref{losse5}) can be illustrated by the the Feynman diagram in Fig.\ref{FigA3}. 
We note that in the inertial system of the right slab the charge density in the left slab, apart from the fluctuations, has an 
additional parallel component of motion, so all parallel coordinates in the left slab have to be 
transformed as in (\ref{njuski}).
Explicitly, the correlation function (\ref{correlajme}) becomes 
\begin{equation}   
S_1({\bf r},{\bf r}_1,t,t_1)=S_1(z,z_1,\brho-{\bf v}t,\brho_1-{\bf v}t_1,t,t_1).
\label{Eka}
\end{equation}
After inserting (\ref{Eka}) into (\ref{losse5}) and the Fourier transformation in parallel coordinates and in time we 
get the formula for energy transfer rate per unit surface area from the left to the right slab 
\begin{equation}
\begin{array}{c}
P_{12}=
-i\hbar\int^{\infty}_{-\infty}dz\int^{\infty}_{-\infty}dz_1\int \frac{d{\bf Q}}{(2\pi)^2}\int^{\infty}_{-\infty}\frac{d\omega}{2\pi}  
\\
\\
\omega'S_1({\bf Q},\left|\omega\right|,z,z_1)V({\bf Q},z,z_3)\otimes 
\\
\\
D_2({\bf Q},\omega',z_3,z_2)
\otimes V({\bf Q},z_2,z_1)
\end{array}
\label{lossaj}
\end{equation}
After using the definitions (\ref{Def1}), (\ref{Def2}) and (\ref{Def3}) equation (\ref{lossaj})  can be written as  
\begin{eqnarray}
P_{12}=
-i\hbar\int \frac{d{\bf Q}}{(2\pi)^2}\int^{\infty}_{-\infty}\frac{d\omega}{2\pi}\  
e^{-2Qa}
S_1({\bf Q},\left|\omega\right|) 
\label{lossos}\\
\omega'
D_2({\bf Q},\omega')
\nonumber
\end{eqnarray}
Using the connection (\ref{impsv}) between the surface correlation function $S$ and the imaginary part of the surface excitation 
propagator R, equation (\ref{lossos}) can be written as
\begin{eqnarray}
P_{12}=
-i\hbar\int \frac{d{\bf Q}}{(2\pi)^2}\int^{\infty}_{-\infty}\frac{d\omega}{2\pi}e^{-2Qa}
\ \omega'[2n_1(\omega)+1] 
\nonumber\\
Im D_1({\bf Q},\omega) D_2({\bf Q},\omega').\hspace{2cm}
\label{loss15}
\end{eqnarray} 
Finally, as the imaginary part of surface excitation propagator (\ref{impsv}) is an odd function 
of frequency, $P_{12}$ given by Eq.(\ref{loss15}) is a real quantity 
\begin{eqnarray}
P_{12}=
\hbar\int \frac{d{\bf Q}}{(2\pi)^2}e^{-2Qa}\int^{\infty}_{-\infty}\frac{d\omega}{2\pi}\  
\omega'[2n_1(\omega)+1] 
\nonumber\\
\nonumber\\
Im D_1({\bf Q},\omega) ImD_2({\bf Q},\omega').\hspace{2cm}
\label{losse16}
\end{eqnarray} 
The Feynman diagram which illustrates equation (\ref{losse16}) is shown in Fig.\ref{FigA3}a. 
We see that if the charge fluctuation is created with the energy $\omega$ it can create 
excitations in the right slab with the energy $\omega'=\omega+{\bf v}{\bf Q}$. This is expected, namely, $\omega$ is the 
energy in the inertial system of the left slab, but in the inertial system of the right slab it 
is Doppler shifted by {\bf v}{\bf Q}. 

In (\ref{losse16}) we have calculated energy transferred from the left to the right slab. However, the part of this 
energy belongs to the quantum mechanical fluctuation which will be reversibly returned back to the left slab. 
We can calculate this part of energy which fluctuates between the slabs by going to the inertial system of the 
left slab and forgeting for the moment the right one. Sitting in the inertial system of the left slab we know that it is in the 
quantumechanical (and thermodynamical) equilibrium with the environment (in this case with the right slab). 
So, the energy just fluctuates between the left slab and the environment, i.e. the energy which is given to the environment is 
exactly equal to the energy which is received from the environment. The energy given to the 
environment, i.e. to the right slab, can be calculated using exactly the same ideas as before, 
except that now the right slab is moving with the velocity $-{\bf v}$ and the left one is at rest. 
Therefore, following the same procedure (\ref{losse4}--\ref{losse16}) 
with the response functions of the right slab transformed as 
\begin{equation}   
D_2({\bf r},{\bf r}_1,t,t_1)=D_2(z,z_1,\brho+{\bf v}t,\brho_1+{\bf v}t_1,t,t_1)
\label{Galtral}
\end{equation}
we obtain the energy that is reversibly given to the right slab 
\begin{eqnarray}
P'_{12}=
\hbar\int \frac{d{\bf Q}}{(2\pi)^2}e^{-2Qa}\int^{\infty}_{-\infty}\frac{d\omega}{2\pi}\ \omega\ [2n_1(\omega)+1]\ 
\nonumber\\
\label{losse17}\\
ImD_1({\bf Q},\omega)ImD_2({\bf Q},\omega').
\nonumber
\end{eqnarray} 
This means that the energy which is irreversibly given to the right slab or dissipated 
energy can be obtained by substracting the reversible contribution $P'_{12}$ from the total energy 
transfer $P_{12}$
\begin{eqnarray}
P_{1}=P_{12}-P'_{12}=\hbar{\bf v}\int \frac{d{\bf Q}}{(2\pi)^2}\ {\bf Q}\ e^{-2Qa}
\nonumber\\
\label{losse18}\\
\int^{\infty}_{-\infty}\frac{d\omega}{2\pi} [2n_1(\omega)+1]\ 
ImD_1({\bf Q},\omega) 
ImD_2({\bf Q},\omega').
\nonumber
\end{eqnarray}
Expression (\ref{losse18}) represents the dissipated power if the charge fluctuation is spontaneously created in the left 
slab. However, the charge fluctuation can also be spontaneously created in the right slab, then the corresponding 
dissipated power can be obtained from (\ref{losse18})  with $1 \leftrightarrow 2$ and  ${\bf v}\leftrightarrow -{\bf v}$.        
Therefore the total dissipated power can be written as 
\begin{eqnarray}
P=P_1+P_2=\hspace{4cm}
\label{loss20}\\
\nonumber\\
\hbar{\bf v}\int \frac{d{\bf Q}}{(2\pi)^2}{\bf Q}e^{-2Qa}\int^{\infty}_{-\infty}
\frac{d\omega}{2\pi}\ sgn(\omega)\hspace{2cm}
\nonumber\\
\nonumber\\
\left\{[2n_1(\omega)+1]ImD_1({\bf Q},\omega)ImD_2({\bf Q},\omega')+\right.
\nonumber\\
\nonumber\\
\left.[2n_2(\omega)+1]ImD_1({\bf Q},\omega')ImD_2({\bf Q},\omega)\right\}.
\nonumber
\end{eqnarray}
This result can be transformed by changing the arguments $\omega+{\bf v}{\bf Q}\rightarrow\omega$ 
and ${\bf Q}\rightarrow-{\bf Q}$ in the second term of (\ref{loss20}) and the frequency integration becomes  
\begin{eqnarray}
2\hbar{\bf v}{\bf Q}\int^{\infty}_{-\infty}\frac{d\omega}{2\pi}\left[n_1(\omega)-n_2(\omega+{\bf v}{\bf Q})\right]\hspace{3cm}
\nonumber\\
\nonumber\\
ImD_1({\bf Q},\omega)ImD_2({\bf Q},\omega+{\bf v}{\bf Q})
\nonumber
\end{eqnarray}
For $T=0$ this reduces to the well known result \cite{trenjePRB}
\begin{equation}
2\hbar{\bf v}{\bf Q}\int^{{\bf Q}{\bf v}}_{0}\frac{d\omega}{2\pi}ImD_1({\bf Q},\omega)ImD_2({\bf Q},{\bf v}{\bf Q}-\omega).\hspace{1cm}
\label{jutro}
\end{equation}

As in the case of van der Waals energy in Sec.\ref{vdWenergyforce} the higher order terms can be included by replacing $v_Q$'s in $D_i$'s  in 
(\ref{jutro}) by an infinite series (\ref{Eka1},\ref{Eka2}), as also shown in Fig.\ref{FigA2}, so that we get
\begin{equation}
P=2\hbar\int\frac{d{\bf Q}}{(2\pi)^2}e^{-2Qa}{\bf v}{\bf Q}
\int^{\infty}_{-\infty}\frac{d\omega}{2\pi}\ C({\bf Q},\omega,\omega')
\end{equation}
where 
\begin{equation}
C({\bf Q},\omega,\omega')=
[n_1(\omega)-n_2(\omega')]\frac{ImD_1({\bf Q},\omega)ImD_2({\bf Q},\omega')}
{\left|1-e^{-2Qa}D_1({\bf Q},\omega)D_2({\bf Q},\omega')\right|^2}
\label{kaka}
\end{equation}
and $\omega'={\bf v}{\bf Q}-\omega$.
Dissipated power can be simply related to the friction force ${\bf F}$ by $P=-{\bf F}{\bf v}$, so that 
\begin{equation}
{\bf F}=-\hbar\int\frac{d{\bf Q}}{(2\pi)^2}e^{-2Qa}{\bf Q}\int^{\infty}_{-\infty}\frac{d\omega}{2\pi}
C({\bf Q},\omega,\omega').
\end{equation}
Obviously, for ${\bf v}\rightarrow 0$ both $P$ and ${\bf F}$ vanish. 
This result agrees with Pendry's alternative derivation \cite{Pendry1,Pendry2}.

The above derivation repeats and generalizes some previously well known results \cite{Pendry1,Persson1,Volokitin1,
Volokitin2,Pendry2,Philbin,coment1,coment2,trenjePRB}. We should note that this derivation takes 
into account not the local but the full microscopically calculated nonlocal response functions 
$R_i({\bf Q},\omega,z,z'); i=1,2$. However, its main purpose is to facilitate the derivation of analogous results for 
the oscillating slabs in Sec.\ref{Sec2}.


\begin{thebibliography}{99}
\bibitem{Laplace} The dielectric substrate screened interaction  ${\tilde v}_Q$ can be obtained by solving Laplace equation $\left\{\frac{\partial^2}{\partial z^2}-Q^2\right\}
\phi_{z'}(z)=4\pi\delta(z-z')$ and applying boundary conditions $D^i_n=D^j_n$ and $\epsilon_iE^i_t=\epsilon_jE^j_t$, where $i\ne j=\pm \Delta/2\pm 0^+$, $\epsilon_{\pm \Delta/2\mp 0^+}=
\epsilon(\omega)$ and $\epsilon_{\pm \Delta/2\pm 0^+}=1$. Finally, ${\tilde v}_{Q}=\phi_{\pm \Delta/2}(\pm \Delta/2)$.
\end{thebibliography}
\end{document}